\documentclass[11pt]{article}
\usepackage{latexsym} \usepackage{epsf}
\usepackage{a4}
\usepackage{amsfonts}
\usepackage{amsmath}
\usepackage{amsthm}
\usepackage{amssymb}

\renewcommand{\theequation}{\thesection.\arabic{equation}}
\newcounter{subequation}[equation]
\makeatletter

\expandafter\let\expandafter\reset@font\csname reset@font\endcsname

\def\subeqnarray{\arraycolsep1pt
    \def\@eqnnum\stepcounter##1{\stepcounter{subequation}%
        {\reset@font\rm(\theequation\alph{subequation})}}
\jot5mm     \eqnarray}

\makeatother

\def\be{\begin{equation}}
\def\ee{\end{equation}}
\def\bea{\begin{eqnarray}}
\def\eea{\end{eqnarray}}
\def\ba{\begin{array}}
\def\ea{\end{array}}
\def\dd{\partial}

\def\half{\frac{1}{2}}

\def\one#1{#1^{\raise5pt\hbox{$\scriptstyle\!\!\!\!1$}}\,{}}
\def\two#1{#1^{\raise5pt\hbox{$\scriptstyle\!\!\!\!2$}}\,{}}

\def\II{\hbox{{1}\kern-.25em\hbox{l}}}
\makeatletter
\def\binrel@#1{\begingroup
  \setboxz@h{\thinmuskip0mu
    \medmuskip\m@ne mu\thickmuskip\@ne mu
    \setbox\tw@\hbox{$#1\m@th$}\kern-\wd\tw@
    ${}#1{}\m@th$}%
  \edef\@tempa{\endgroup\let\noexpand\binrel@@
    \ifdim\wdz@<\z@ \mathbin
    \else\ifdim\wdz@>\z@ \mathrel
    \else \relax\fi\fi}%
  \@tempa
}
\let\binrel@@\relax
\def\overset#1#2{\binrel@{#2}%
  \binrel@@{\mathop{\kern\z@#2}\limits^{#1}}}
\def\underset#1#2{\binrel@{#2}%
  \binrel@@{\mathop{\kern\z@#2}\limits_{#1}}}
\makeatother
\newfont{\bbd}{msbm10 scaled\magstep1}
\def\C{\hbox{\bbd C}}

\def\V{\hbox{\bbd V}}

\def\RR{{\mathcal R}}

\def\RR{{\mathcal R}}

\newtheorem{prop}{Proposition}

\begin{document}

{\begin{center}
{\LARGE {Yang-Baxter $R$ operators and parameter permutations
} } \\ [8mm]
{\large  S. Derkachov$^{a}$
\footnote{e-mail:derkach@euclid.pdmi.ras.ru},
D. Karakhanyan$^b$\footnote{e-mail: karakhan@lx2.yerphi.am} \&
R. Kirschner$^c$\footnote{e-mail:Roland.Kirschner@itp.uni-leipzig.de} \\
[3mm] }
\end{center}

\begin{itemize}
\item[$^a$]
St. Petersburg Department of Steklov Mathematical Institute
of Russian Academy of Sciences,
Fontanka 27, 191023 St. Petersburg, Russia.
\item[$^b$]
Yerevan Physics Institute, \\
Br. Alikhanian st. 2, 375036 Yerevan, Armenia.
\item[$^c$]
 Institut f\"ur Theoretische
Physik, Universit\"at Leipzig, \\
PF 100 920, D-04009 Leipzig, Germany
\end{itemize}

\vspace{3cm} \noindent {\bf Abstract} 
 We present an uniform construction of the solution to the Yang-
Baxter equation with the symmetry algebra $s\ell(2)$ and its
deformations: the q-deformation and the elliptic deformation or
Sklyanin algebra. The R-operator acting in the tensor product of two
representations of the symmetry algebra with arbitrary spins
$\ell_1$ and $\ell_2$ is built in terms of products of three basic
operators $\mathcal{S}_1, \mathcal{S}_2,\mathcal{S}_3$ which are
constructed explicitly. They have the simple meaning of representing
elementary permutations of the symmetric group $\mathfrak{S}_4$, the
permutation group of the four parameters entering the RLL-relation.

\renewcommand{\refname}{References.}
\renewcommand{\thefootnote}{\arabic{footnote}}
\setcounter{footnote}{0}

\newpage

\section{Introduction}
\setcounter{equation}{0}

The Yang-Baxter equation and its solutions play a key role in the
theory of the completely integrable quantum models~
\cite{Baxter,FTa,KS,Jimbo,Faddeev}. 
The general solution of the Yang-Baxter
equation (R-matrix) is the operator $\mathcal{R}(u)$ acting in a
tensor product $\mathbf{V}_1\otimes\mathbf{V}_2$ of two linear
spaces. There exists the natural construction of the solution to the
Yang-Baxter equation. The solution of the problem consists of two
steps:
\begin{itemize}
\item on the first stage we construct the solution to so called
$RLL$-relation
\item on the second stage we prove that operator
$\mathcal{R}_{12}(u-v)$ constructed as the solution of the
$RLL$-relation obeys the general Yang-Baxter equation.
\end{itemize}
 Choosing one of the three spaces involved to carry a
fundamental representation of the symmetry algebra the Yang-Baxter
equation is reduced to the simpler defining equation for the
R-matrix~\cite{KRS}
 \be\label{rll}
\mathcal{R}_{12}(u-v)\,\mathrm{L}_1(u)\,\mathrm{L}_2(v)=
\mathrm{L}_2(v)\,\mathrm{L}_1(u)\,\mathcal{R}_{12}(u-v)\ ,
 \ee
where $\mathrm{L}(u)$ is the Lax matrix. In the cases of the
symmetry algebra $s\ell(2)$ and its q-deformation the R-matrix can
be obtained by the following method~\cite{Faddeev,KRS}: The R-matrix is
a function of the Casimir operator and the defining RLL-equation is
reduced to a recurrence relation for the function of one variable.

We shall give an uniform construction of the solution of the
RLL-relation extending to the three cases of the symmetry algebra
$s\ell(2)$ and its two deformations, the q-deformation and the
elliptic deformation or Sklyanin algebra~\cite{Sklyanin}.
It is well known that the resulting R-matrices are building blocks in 
the construction of integrable spin chain models. Whereas in questions
related to condensed matter physics the sites of the chain
carry usually finite-dimensional representations the case of
infinite-dimensional representations turned out to be relevant in studies
of QCD and super Yang-Mills theories, in particular in calculating the
anomalous dimensions of composite operators or of the Regge 
singularities of multi-gluon exchange (cf. \cite{BBGK}). 
It is not known so far whether the deformed symmetry cases are relevant 
in the Yang-Mills context. The explicit formulation of the results 
in analogous form for all cases may be useful to find new applications.

The main idea of our construction is quite simple: 
The R-matrix represents some 
 particular element of the group $\mathfrak{S}_4$ of permutations
of four parameters entering the RLL-relation and it can be
constructed from the simple building blocks, the operators
$\mathcal{S}_1, \mathcal{S}_2,\mathcal{S}_3$  corresponding to the
elementary permutations. The Lax matrix $\mathrm{L}(u)$ depends on
two parameters: the spin of representation $\ell$ and the spectral
parameter~$u$. It is useful to introduce  two related parameters
$u_1$ and $u_2$,  $\ u_1 = \frac{u}{2\eta} + \ell\ ;\ u_2 =
\frac{u}{2\eta} -1 -\ell $,  where $\eta$ is a free parameter in the
case of the Sklyanin algebra. In the cases of no deformation and
q-deformation it can be fixed as $\eta = \half$ without loss of
generality. After extracting the operator of permutation
$\mathcal{P}_{12}$ from the R-matrix
$\RR_{12}=\mathcal{P}_{12}\check{\RR}_{12}$ the defining equation is
rewritten in "check formalism",
 \be\label{RLL}
\check{\RR}_{12}\cdot \mathrm{L}_1(u_1,u_2)\mathrm{L}_2(v_1,v_2) =
\mathrm{L}_1(v_1,v_2)\mathrm{L}_2(u_1,u_2)\cdot \check{\RR}_{12}.
 \ee
This equation admits the natural interpretation: The operator
$\check{\mathcal{R}}_{12}$ interchanges the set of parameters
$(u_1,u_2)$ of the first Lax-matrix with the set of parameters
$(v_1,v_2)$ of the second Lax-matrix. It corresponds to the special
permutation $s$ in the group of permutations $\mathfrak{S}_4$ of
four parameters,
$$
(\overset{s}{\overbrace{\underline{u_1,u_2},\underline{v_1,v_2}}})\mapsto
(\underline{v_1,v_2},\underline{u_1,u_2}).
$$
 Arbitrary permutations in the group $\mathfrak{S}_4$ can be
constructed from the elementary transpositions $\mathrm{s}_{1}$,
$\mathrm{s}_{2}$ and $\mathrm{s}_{3}$ which interchange only two
nearest neighboring components in the set $(u_1,u_2,v_1,v_2)$. We
look for  the operators $\mathcal{S}_i$ representing these
elementary transpositions
$$
(\overset{s_1}{\overbrace{u_1\ ,\ u_2}},
\overset{s_3}{\overbrace{v_1\ ,\ v_2}})\ \ ;\ \
\mathcal{S}_1\,\mathrm{L}_1(\overset{s_1}{\overbrace{u_1,u_2}}) =
\mathrm{L}_1(u_2,u_1)\mathcal{S}_1\ ,\
\mathcal{S}_3\,\mathrm{L}_2(\overset{s_3}{\overbrace{v_1,v_2}}) =
\mathrm{L}_2(v_2,v_1)\mathcal{S}_3\ ,
$$
$$
(u_1\ ,\overset{s_2}{\overbrace{u_2,v_1}},\ v_2)\ \ ;\ \
\mathcal{S}_2\,\mathrm{L}_1(u_1,
\overset{s_2}{\overbrace{u_2)\,\mathrm{L}_2(v_1}},v_2)=
\mathrm{L}_1(u_1,v_1)\,\mathrm{L}_2(u_2,v_2)\mathcal{S}_2\,.
$$
These equations appear to be much simpler than the initial defining
equation for the R-operator and the solution can be obtained in a
closed form. Finally we construct the R-matrix as the composite
object out of the simplest building blocks,  the operators
$\mathcal{S}_1,\mathcal{S}_2,\mathcal{S}_3$.

Note that there are different possible 
levels of resolution into more elementary permutations:
\begin{itemize}
 \item
 On the first level we have the group of permutation
$\mathfrak{S}_2$ of two pairs of parameters $(u_1,u_2)$ and
$(v_1,v_2)$ and the R-matrix is the elementary operator
corresponding to this permutation.
 \item
 On the second level we allow the separate permutations $u_1
\leftrightarrow v_1$ or $u_2\leftrightarrow v_2$ so that our group
of permutations is $\mathfrak{S}_2\times\mathfrak{S}_2$. On this
level the operators $\check{\mathcal{R}}^{(1)}$ and $\check{
\mathcal{R}}^{(2)}$ corresponding to these permutations are the
elementary building blocks but the R-matrix is a composite object.
 \item
 On the third level we allow any permutations of parameters so
that our group of permutations becomes $\mathfrak{S}_4$. On this
level the operators $\mathcal{S}_{i}$, $i=1,2,3$ are elementary
building blocks. Now the operators $\check{\mathcal{R}}^{(1)}$ ,
$\check{\mathcal{R}}^{(2)}$ and R-matrix are composite objects. So
we have the chain of inclusions with  increasing symmetry
$$
\mathfrak{S}_2 \rightarrow \mathfrak{S}_2\times
\mathfrak{S}_2\rightarrow\mathfrak{S}_4.
$$
\end{itemize}

Thus the R-matrix is factorized into operators representing more
elementary permutations. We shall construct the elementary
operator factors $\mathcal{S}_{i}$, $i=1,2,3$ and proof that
they generate a representation of the permutation group for generic
values of the representation parameters $u_1, u_2, v_1, v_2$.
Then the Yang-Baxter relation for the general R-matrix composed out
of these factors is just a consequence of the permutation
group relations.

The solution of the Yang-Baxter relation by factorization into
elementary parameter permutations has been given previously in
\cite{DM} for the symmetry $s\ell(N)$~. Here we show that the method
extends to the deformations of $sl(2)$. The construction of
$\mathcal{S}_{i}$ and the permutation group proofs will be done
uniformly for the cases of no deformation and of the quantum and
elliptic deformations emphasizing the analogies. Guided by the
analogies we are able to present a relatively simple solution also
for the non-trivial elliptic case.

For generic values of the spins $\ell_1, \ell_2$ the operators
$\check{\mathcal{R}}^{(i)}$~\cite{D,DM1,DKK} act within the tensor
product of the corresponding representation spaces, whereas the
operators $\mathcal{S}_{i}$ map to tensor products with changed spin
values. Moreover, in the case of finite dimensional representations
only the R-matrix itself leaves the tensor product space invariant.
At some special values of the representation parameters the
elementary operators $\mathcal{S}_{i}$ become singular on  parts of
their spectra or develop  kernels. Then the permutation group
properties become invalid. In this paper we restrict ourselves to
generic parameter values and postpone those more subtle questions
related to representation theory to the next paper. There we plan
also to consider the uniform construction of Baxter $Q$ operators in
this approach.

The presentation is organized as follows. In Section 2 we give the
uniform expression for the fundamental R-matrices and Lax matrices
for the three cases: no deformation, q-deformation and elliptic
deformation of the symmetry algebra $s\ell(2)$. In Section 3 we
discuss the connection between RLL-relations and the group of
permutations $\mathfrak{S}_4$. In Section 4 we construct the
operators $\mathcal{S}_1,\mathcal{S}_2,\mathcal{S}_3$ representing
the elementary transpositions and in Section 5 we prove the
Coxeter relations for them. In Section 6 we construct the R-matrix
and prove that it obeys the Yang-Baxter relation. In Section 7 we
construct the operators $\check{\RR}^{(1)}$ and
$\check{\RR}^{(2)}$ and prove the corresponding Yang-Baxter
relations for them. Finally, in Section 8 we summarize.

\section{Yang Baxter $\RR$ matrices related to $s\ell_2$}
\setcounter{equation}{0}

The Yang-Baxter relation, \be \label{YB} \mathcal{R}_{12} (u-v)
\mathcal{R}_{13}(u) \mathcal{R}_{23}(v) =\mathcal{R}_{23}(v)
\mathcal{R}_{13}(u)\mathcal{R}_{12}(u-v), \ee of operators acting
on the tensor product $\V_1\otimes \V_2 \otimes \V_3$ where
$\RR_{ij} $ is acting non-trivially only on $\V_i, \V_j $, has the
three well known solutions in the case $\V_i \equiv {\C}^2 $
related to $s\ell_2$
 \be \label{fund}
\RR_{12} (u) = \half \sum_{a=0}^3 w_{a} (u+ \eta) \sigma^a \otimes
\sigma^a.
 \ee
Here $\sigma^a$ denote the Pauli matrices, $\sigma^0 = \II$. There
are the three cases of no deformation , quantum deformation and
elliptic deformation . The parameter $\eta$ can be fixed in the
first two cases, $2 \eta = 1$, but is variable in the last case. The
weight functions are depending on case,
\begin{itemize}
\item no deformation
 \be\label{I}
w_{0}(u) = 2 u;\qquad  w_{1} (u) = w_{2}(u) = w_{3}(u) = 1,
 \ee
 \item
quantum deformation 
 \be\label{II}
w_{0}(u) =  \frac{q^u-q^{-u}}{q^{\half}-q^{-\half}}; \qquad
w_{3}(u) = \frac{q^u+q^{-u}}{q^{\half}+q^{-\half}};\qquad w_{1}(u) =
w_{2}(u) = 1,
 \ee
\item elliptic deformation
 \be \label{III}
w_{a}(u) = {\theta_{a+1} (u|\tau) \over \theta_{a+1}(\eta|\tau) }.
 \ee
\end{itemize}
The  relation (\ref{YB}) is the source of a rich algebraic structure
and integrable quantum systems. In the case $\V_1 ={\C}^2$ and
$\V_2$ arbitrary one calls $\RR_{12} (u+\eta) = \mathrm{L}(u)$ Lax
matrix,
 \be \label{Lax}
 \mathrm{L}(u) = \half \sum_{a=0}^3 w_{a} (u) \sigma^a
\otimes \mathbf{S}^a = \half \left(
\begin{array}{cc}
w_0(u)\mathbf{S}^0+w_3(u)\mathbf{S}^3 &
w_1(u)\mathbf{S}^1+i w_2(u)\mathbf{S}^2 \\
w_1(u)\mathbf{S}^1-i w_2(u)\mathbf{S}^2&
w_0(u)\mathbf{S}^0-w_3(u)\mathbf{S}^3
\end{array} \right ),
 \ee
where the operators $\mathbf{S}^a$ generate the universal enveloping
algebra $\mathrm{U}(s\ell_2)$ in the first case, its quantum
deformation $\mathrm{U}_q(s\ell_2)$ (deformation parameter $q$) in
the second case and its elliptic deformation(parameters $\tau,
\eta$), Sklyanin algebra, in the third case. In the first two cases
$\mathbf{S}^0, \mathbf{S}^3$ are not independent and they are
related to the generators in conventional notation
\begin{itemize}
\item no deformation
$$
\mathbf{S}^0 = \II, \qquad \mathbf{S}^3 = 2 S , \qquad \mathbf{S}^1
\pm i \mathbf{S}^2 = 2 S^{\pm}
$$
where the conventional generators for the representation with spin $\ell$ are
 \be \label{sl2}
 S^- = - \dd, \qquad S = z \dd - \ell, \qquad S^+ = z^2 \dd-2\ell z ,
 \ee
After substitution in the Lax matrix~(\ref{Lax}) the
functions~(\ref{I}) and generators in terms of $z, \dd$ as in
(\ref{sl2}) we obtain that the Lax matrix decomposes into factors
depending on one of the two elementary operators $z$ or $\dd$ only.
The spectral parameter $u$ and spin $\ell$ enter in the combinations
$u_1 = u+ \ell$ and $u_2 = u-1 -\ell$
$$
 \mathrm{L} (u_1,u_2) = \left(
\begin{array}{cc}
1 &0 \\
z & u_1
\end{array} \right )\left(
\begin{array}{cc}
1 &- \dd \\
0 & 1
\end{array} \right )\left(
\begin{array}{cc}
u_2 &0 \\
-z & 1
\end{array} \right ).
$$
\item quantum deformation
 \be \label{S03}
\mathbf{S}^0 = \frac{q^{S} +q^{-S}}{q^{\half} + q^{-\half}}\ ;\qquad
\mathbf{S}^3 = \frac{q^{S} - q^{-S}}{q^{\half} - q^{-\half}}\ ;\
\qquad \mathbf{S}^1 \pm i \mathbf{S}^2 = 2 S^{\pm}.
 \ee
The conventional generators for the representation with spin $\ell$
are
 \be \label{sl2q}
S^- = - \frac{1}{z} [z \dd]_q, \qquad S = z \dd - \ell, \qquad S^+ =
z [z\dd -2\ell]_q,
 \ee
where $[x]_q$ is the usual notation for q-numbers $[x]_q = {q^x -
q^{-x} \over q-q^{-1} }$. Substituting the functions~(\ref{II}) and
$q$ deformed generators $S^a$  from (\ref{sl2q}) in the Lax matrix
(\ref{Lax}) we obtain a factorized form with features analogous to
the undeformed case~\cite{KZ,Zabrodin,DKK}
$$
\mathrm{L}(u_1,u_2) = \left(
\begin{array}{cc}
1 & 1 \\
z q^{-u_1} & z q^{u_1}
\end{array} \right )\left(
\begin{array}{cc}
q^{z \dd +1} & 0 \\
0 & q^{- z \dd -1}
\end{array} \right )\left(
\begin{array}{cc}
q^{u_2} & -z^{-1} \\
- q^{-u_2} & z^{-1}
\end{array} \right )
$$
\item the generators $\mathbf{S}^a$ of the Sklyanin algebra can be
expressed in terms of $z, \dd$ as follows
 \be \label{Se}
\mathbf{S}^a =\frac{(i)^{\delta_{a,2}}
\theta_{a+1}(\eta)}{\theta_1(2\eta z) } \left( \theta_{a+1}
\left[2\eta(z-\ell)\right]\cdot \mathrm{e}^{\dd} - \theta_{a+1}
\left[-2\eta(z-\ell)\right]\cdot \mathrm{e}^{-\dd} \right).
 \ee

Note that in order to simplify the formulae in the elliptic case we
change the variable $z \to \eta z$ in comparison to the standard
notations~\cite{Sklyanin,Zabrodin} and absorb the model parameter
$\eta$ which indeed plays important role in XYZ model in contrast
with simpler XXX and XXZ case, but in present context we do not
touch the subtle questions related to the elliptic curve
parametrizing the model under consideration. The needed properties
of the $\theta$-~functions are listed in the Appendix. Also in this
case an analogous factorization of the Lax matrix can be written
after substituting in (\ref{Lax}) the generators in terms of $z,
\dd$~\cite{KZ,Zabrodin}
$$
\mathrm{L}(u_1,u_2)= \frac{1}{\theta_1(2\eta z)} \cdot \left(\!\!\!
\begin{array}{cc}
\left(z-u_1\right)_3 & -\left(z+u_1\right)_3 \\
-\left(z-u_1\right)_4 & \left(z+u_1\right)_4
\end{array}\!\!\! \right )\cdot
\left(\!\!\!
\begin{array}{cc}
\mathrm{e}^{\dd} &0\\
0 & \mathrm{e}^{-\dd }
\end{array}\!\!\! \right )\left(\!\!\!
\begin{array}{cc}
\left(z+u_2\right)_4 & \left(z+u_2\right)_3 \\
\left(z-u_2\right)_4 & \left(z-u_2\right)_3
\end{array}\!\!\! \right )
$$
where $ u_1 = \frac{u}{2\eta} + \ell\ ;\ \ \ u_2 = \frac{u}{2\eta}
- \ell -1 $ and for simplicity we use the notation
$$\left(x \right)_3 = \theta_3\left(\eta
x|\frac{\tau}{2}\right)\ ;\ \ \left(x \right)_4 =
\theta_4\left(\eta x|\frac{\tau}{2}\right).
$$
\end{itemize}
We obtain the uniform expressions of Lax factorization.

\begin{prop}
In terms of a Heisenberg pair $z, \dd$ the Lax matrices
(\ref{Lax}) can be written in factorized form as \be \label{Lfact}
\mathrm{L}_c (u_1,u_2)= [u_1]_c \ \mathrm{V}_c^{-1} (u_1,z) \
\mathrm{D}_c \ \mathrm{V}_c(u_2,z) \ee where the subscript $c =
o,q,e$ labels the three cases of no deformation (o), quantum
deformation (q) and elliptic deformation (e). The first factor is
a usual number $[u]_o = u$, q-number $[u]_q = {q^u - q^{-u} \over
q-q^{-1} }$ or elliptic number $[u]_e = \theta_1(2u\eta)$ and the
others are $2 \times 2$ matrices. The matrices $\mathrm{D}_c$ do
not depend on the parameters $u_1, u_2$, $ u_1 = \frac{u}{2\eta_c}
+ \ell\ ;\ u_2 = \frac{u}{2\eta_c} -1 -\ell
$, with $2\eta_o = 2\eta_q = 1, 2\eta_e = 2\eta $ and are given in
terms of operator of differentiation $\dd$ (o), dilatation by
$q^{\pm1}$ (q) or shift by $\pm 1$ (e)
\bea \label{D} \mathrm{D}_o
= \left(
\begin{array}{cc}
1 &- \dd \\
0 & 1
\end{array} \right )\ ;\ \mathrm{D}_q = \left(
\begin{array}{cc}
q^{z \dd +1} & 0 \\
0 & q^{- z \dd -1}
\end{array} \right )\ ;\ \mathrm{D}_e = \left(
\begin{array}{cc}
\mathrm{e}^{\dd} &0\\
0 & \mathrm{e}^{- \dd }
\end{array} \right ).
\nonumber
\eea
The matrices $V_c$ depend on the representation parameters
and on $z$
 \bea \label{V} \mathrm{V}_o(u,z) = \left(\!\!\!
\begin{array}{cc}
u &0 \\
-z & 1
\end{array}\!\!\! \right );\;\;
\mathrm{V}_q(u,z) = \left(\!\!\!
\begin{array}{cc}
q^u & -z^{-1} \\
- q^{-u} & z^{-1}
\end{array}\!\!\! \right );\;\;
\mathrm{V}_e(u,z) = \left(\!\!\!
\begin{array}{cc}
\left(z+u\right)_4 & \left(z+u\right)_3 \\
\left(z-u\right)_4 & \left(z-u\right)_3
\end{array}\!\!\! \right )
\nonumber \eea
\end{prop}

\section{The RLL-relation and the permutation group}
\setcounter{equation}{0}

Let us consider the reduction of the general Yang-Baxter
equation~(\ref{YB}) in the special case when $\V_3=\C^2$. The
operator $\RR_{13}(u)$ is reduced to the Lax matrix
$\mathrm{L}_1(u)$, operator $\RR_{23}(v)$ is reduced to
$\mathrm{L}_2(v)$ so that one obtains the defining RLL-relation for
the operator $\RR_{12}(u-v)$ (\ref{rll}). It is useful to extract
the permutation operator from the $\RR$-matrix
$$\RR_{12}=\mathcal{P}_{12}\,\check{\mathcal{R}}_{12} , $$
where $\mathcal{P}_{12}\,\Phi(z_1,z_2)=\Phi(z_2,z_1)$. For the
operator $\check{\mathcal{R}}_{12}$ the $RLL$-relation takes the
form (\ref{RLL}) where $\mathrm{L}_1$ is to be substituted as in
(\ref{Lfact}) with $z,\dd$ replaced by $z_1, \dd_1$ and
$\mathrm{L}_2$ with $z,\dd$ replaced by $z_2, \dd_2$. The parameters
are $u_1 = \frac{u}{2\eta_c}+ \ell_1, u_2 =
\frac{u}{2\eta_c}-1-\ell_1$ and $v_1 = \frac{v}{2\eta_c} + \ell_2,
v_2 = \frac{v}{2\eta_c} -1 -\ell_2$. In equation~(\ref{RLL})
$\check{\mathcal{R}}_{12}(u-v)$  is some operator depending on the
Heisenberg pairs $z_1, \dd_1$ and $z_2, \dd_2$.

As it is mentioned in Introduction the operator
$\check{\mathcal{R}}_{12}(u-v)$ in (\ref{RLL}) corresponds to the
special permutation $s$ in the group of permutations
$\mathfrak{S}_4$ of four parameters 
~$\mathbf{u} = (u_1,u_2,v_1,v_2)$.
$$
s \rightarrow \check{\mathcal{R}}_{12}(u-v) \ ;\
s(u_1,u_2,v_1,v_2)=(v_1,v_2,u_1,u_2).
$$
The arbitrary permutation from the group $\mathfrak{S}_4$ can be
constructed from the elementary transpositions $\mathrm{s}_{1}$,
$\mathrm{s}_{2}$ and $\mathrm{s}_{3}$
$$
\mathrm{s}_{1}\mathbf{u} = (u_2,u_1,v_1,v_2)\ ;\
\mathrm{s}_{2}\mathbf{u}
 = (u_1,v_1,u_1,v_2) \ ;\
\mathrm{s}_{3}\mathbf{u} = (u_1,u_2,v_2,v_1)
$$
which interchange only two nearest neighboring components in the set
$\mathbf{u}= (u_1,u_2,v_1,v_2)$. 
For example the permutation $s$ has the
following decomposition $s = s_2 s_1s_3s_2$. It is natural to search
the operators $\mathcal{S}_i(u_1,u_2,v_1,v_2) =
\mathcal{S}_i(\mathbf{u})$ representing these elementary
transpositions
$$
(\overset{\mathcal{S}_1}{\overbrace{u_1\ ,\ u_2}},
\overset{\mathcal{S}_{3}}{\overbrace{v_1\ ,\ v_2}})\ ;\ (u_1\
,\overset{\mathcal{S}_2}{\overbrace{u_2,v_1}},\ v_2).
$$
These operators obey the following defining equations
\begin{equation}\label{RLL13}
\mathcal{S}_1(\mathbf{u})\,\mathrm{L}_1(u_1,u_2) =
\mathrm{L}_1(u_2,u_1)\mathcal{S}_1(\mathbf{u})\ ;\
\mathcal{S}_3(\mathbf{u})\,\mathrm{L}_2(v_1,v_2) =
\mathrm{L}_2(v_2,v_1)\mathcal{S}_3(\mathbf{u})
\end{equation}
\begin{equation}\label{RLL2}
\mathcal{S}_2(\mathbf{u})\,\mathrm{L}_1(u_1,u_2)\,\mathrm{L}_2(v_1,v_2)=
\mathrm{L}_1(u_1,v_1)\,\mathrm{L}_2(u_2,v_2)\mathcal{S}_2(\mathbf{u})\,.
\end{equation}
and our first step will be the explicit construction of these
operators in  Section~\ref{constr}. Having these operators we can
construct the R-matrix.
\begin{prop} The operator $\check{\mathcal{R}}_{12}$
\be\label{R} \check{\mathcal{R}}_{12}(u-v) =
\mathcal{S}_2(s_1s_3s_2\mathbf{u})
\mathcal{S}_1(s_3s_2\mathbf{u})\mathcal{S}_3(s_2\mathbf{u})
\mathcal{S}_2(\mathbf{u}) \ee obeys the relation~(\ref{RLL})
provided the operators $\mathcal{S}_i$ obey the
relations~(\ref{RLL13}),~(\ref{RLL2}).
\end{prop}
This expression for the R-matrix corresponds to the particular
decomposition of the permutation $s$: $s = s_2 s_1s_3s_2$. In the
last section we shall present equivalent expressions corresponding
to another decompositions. We shall see that the operators
$\mathcal{S}_i$ have the special dependence on parameters
\begin{equation}\label{depend}
\mathcal{S}_1(\mathbf{u}) = \mathcal{S}_1(u_1-u_2)\ ;\
\mathcal{S}_2(\mathbf{u}) = \mathcal{S}_2(u_2-v_1)\ ;\
\mathcal{S}_3(\mathbf{u}) = \mathcal{S}_3(v_1-v_2)
\end{equation}
so that the operator $\check{\mathcal{R}}_{12}(u-v)$ depends on
the difference of spectral parameters as it should be.  We have
the correspondence
\begin{equation}
s_i \rightarrow \mathcal{S}_i(\mathbf{u})\ ;\ s_i s_j \rightarrow
\mathcal{S}_i(s_j\mathbf{u})\mathcal{S}_j(\mathbf{u})
\end{equation}
and to prove that we obtain the representation of the permutation
group $\mathfrak{S}_4$ it remains to prove the corresponding
defining (Coxeter) relations for the generators
\begin{equation}\label{def1}
s_i s_i = \II \rightarrow
\mathcal{S}_i(s_i\mathbf{u})\mathcal{S}_i(\mathbf{u})= \II\ ;\
s_1s_3 = s_3s_1 \rightarrow
\mathcal{S}_1(s_3\mathbf{u})\mathcal{S}_3(\mathbf{u})=
\mathcal{S}_3(s_1\mathbf{u})\mathcal{S}_1(\mathbf{u})
\end{equation}
\begin{equation}\label{def2}
s_1 s_2 s_1 = s_2 s_1 s_2 \rightarrow
\mathcal{S}_1(s_2s_1\mathbf{u})\mathcal{S}_2(s_1\mathbf{u})
\mathcal{S}_1(\mathbf{u})=
\mathcal{S}_2(s_1s_2\mathbf{u})\mathcal{S}_1(s_2\mathbf{u})
\mathcal{S}_2(\mathbf{u})
\end{equation}
\begin{equation}\label{def3}
s_2 s_3 s_2 = s_3 s_2 s_3 \rightarrow
\mathcal{S}_2(s_3s_2\mathbf{u})\mathcal{S}_3(s_2\mathbf{u})
\mathcal{S}_2(\mathbf{u})=
\mathcal{S}_3(s_2s_3\mathbf{u})\mathcal{S}_2(s_3\mathbf{u})
\mathcal{S}_3(\mathbf{u})
\end{equation}
In  Section~\ref{proof} we prove that the obtained operators
$\mathcal{S}_i(\mathbf{u})$ obey these defining relations. After
this we shall prove that Yang-Baxter relation~(\ref{YB}) for the
operator~(\ref{R}) is consequence of the Coxeter
relations~(\ref{def1}),~(\ref{def2}) and~(\ref{def3}).

\section{Elementary permutation operators $\mathcal{S}_1$,
$\mathcal{S}_2$ and $\mathcal{S}_3$} \label{constr}
\setcounter{equation}{0}

\subsection{The operator $\mathcal{S}_2$}

Consider the defining condition for $\mathcal{S}_2$ (\ref{RLL2})
where $\mathrm{L}_1$ is to be substituted as in (\ref{Lfact}) with
$z,\dd$ replaced by $z_1, \dd_1$ and $\mathrm{L}_2$ with $z,\dd$
replaced by $z_2, \dd_2$
$$ \mathcal{S}_2\cdot [u_1] \ \mathrm{V}^{-1}
(u_1,z_1) \ \mathrm{D}_1 \ \mathrm{V}(u_2,z_1)\cdot [v_1] \
\mathrm{V}^{-1} (v_1,z_2) \ \mathrm{D}_2\ \mathrm{V}(v_2,z_2) =
$$
$$ = [u_1] \
\mathrm{V}^{-1} (u_1,z_1) \ \mathrm{D}_1\ \mathrm{V}(v_1,z_1)
\cdot [u_2] \ \mathrm{V}^{-1} (u_2,z_2) \ \mathrm{D}_2\
\mathrm{V}(v_2,z_2)\cdot \mathcal{S}_2 $$ This equation suggests
the ansatz $ \mathcal{S}_2 = \mathcal{S}_2(z_1,z_2) $ as a
multiplication operator independent of $\dd_1, \dd_2$. Then the
operator $\mathcal{S}_2$ commutes with the matrices $\mathrm{V}^{-1}
(u_1,z_1)$ and $\mathrm{V}(v_2,z_2)$ so that they can be cancelled
and we immediately obtain a much simpler defining equation for
the function $\mathcal{S}_2(z_1,z_2)$
 \be\label{eq} [v_1]\cdot
\mathrm{D}_1^{-1}\mathcal{S}_2(z_1,z_2)\ \mathrm{D}_1 \ \cdot
\mathrm{V}(u_2,z_1)\ \mathrm{V}^{-1} (v_1,z_2) = [u_2] \cdot
\mathrm{V}(v_1,z_1)\ \mathrm{V}^{-1} (u_2,z_2) \cdot \mathrm{D}_2\
\mathcal{S}_2(z_1,z_2) \mathrm{D}_2^{-1}.
 \ee
It remains to solve this equation in each case.

\begin{itemize}

\item In the case of no deformation the defining
condition~(\ref{eq}) results in
$$
\left(
\begin{array}{cc}
1 & \dd_1 \ln \mathcal{S}_2\\
0 & 1
\end{array} \right )
\left(
\begin{array}{cc}
u_2 & 0 \\
z_2-z_1 & v_1
\end{array} \right )
=
\left(
\begin{array}{cc}
v_1 &0 \\
z_2-z_1 & u_2
\end{array} \right )
\left(
\begin{array}{cc}
1 &- \dd_2 \ln \mathcal{S}_2 \\
0 & 1
\end{array} \right )
$$
The wanted function of $z_1, z_2$ has to obey two relations,
$$ \dd_1 \mathcal{S}_2(z_1,z_2) = - \dd_2 \mathcal{S}_2(z_1,z_2)\ ; \ \ \
(z_2-z_1) \dd_2 \ln\mathcal{S}_2(z_1,z_2) = u_2 - v_1 $$ This
results in the permutation operator $\mathcal{S}_2$ up to
normalization \be \label{S2o} \mathcal{S}_2(z_1,z_2) =
(z_2-z_1)^{u_2 - v_1}. \ee

\item In the case of quantum deformation the defining
condition~(\ref{eq}) in matrix form is
$$
\left(
\begin{array}{cc}
 \mathcal{S}_2 (q^{-1} z_1,z_2)& 0\\
0 &  \mathcal{S}_2 (qz_1, z_2)
\end{array} \right )
\left(
\begin{array}{cc}
q^{u_2} -\frac{z_2}{z_1} q^{-v_1} & q^{u_2} -\frac{z_2}{z_1} q^{v_1} \\
-q^{-u_2} +\frac{z_2}{z_1} q^{-v_1} & -q^{-u_2} +\frac{z_2}{z_1} q^{v_1}
\end{array} \right )
= $$ $$ =\left(
\begin{array}{cc}
q^{v_1} -\frac{z_2}{z_1} q^{-u_2} & q^{v_1} -\frac{z_2}{z_1} q^{u_2} \\
-q^{-v_1} +\frac{z_2}{z_1} q^{-u_2} & -q^{-v_1} +\frac{z_2}{z_1} q^{u_2}
\end{array} \right )
\left(
\begin{array}{cc}
 \mathcal{S}_2(z_1,qz_2) &0 \\
0 &  \mathcal{S}_2(z_1, q^{-1} z_2)
\end{array} \right ).
$$
Equating the diagonal elements of the matrices resulting on both sides
leads to the conclusion
$$ \mathcal{S}_2 (z_1, z_2) = z_1^{u_2-v_1} \cdot
\Phi \left(\frac{z_2}{z_1}\right). $$
The resulting equations from the two off-diagonal elements are
compatible and lead to the difference equation
$$\Phi(qx) (1-x q^{v_1-u_2} ) = \Phi (q^{-1}x) (1-x q^{u_2-v_1}) $$
Therefore, in the $q$ deformed case the permutation operator
$\mathcal{S}_2$ has the form \be \label{S2q} \mathcal{S}_2 =
z_{1}^{u_2 - v_1} \ {(\frac{z_2}{z_1}q^{1-u_2 +v_1}; q^2 )
\over
(\frac{z_2}{z_1} q^{1+ u_2 - v_1}; q^2) }. \ee
We use the standard
notation $ (x; q^2) = \prod_{k=0}^{\infty} (1- x q^{2k}) $. In
\cite{KKM} this expression appeared as the appropriate
generalization to the $q$ deformed case of $(z_2-z_1)^{u_2-v_1}$
(\ref{S2o}) in the role of  conformal propagators.

\item In the case of elliptic deformation the defining
equation~(\ref{eq}) results in the systems of four difference
equations

\end{itemize}
$$
\theta(z_1+z_2+u_2-v_1)\theta(z_1-z_2+u_2 +v_1)
\mathcal{S}_2(z_1-1,z_2)\!=\!\theta(z_1+z_2
+v_1-u_2)\theta(z_1-z_2+v_1+u_1)\mathcal{S}_2 (z_1,z_2+1),
$$
$$
\theta(z_1+z_2-u_2+v_1)\theta(z_1-z_2-u_2- v_1)
\mathcal{S}_2 (z_1+1,z_2)\!=\!\theta (z_1+z_2
-v_1-u_2)\theta(z_1-z_2-v_2-u_2)\mathcal{S}_2 (z_1,z_2-1),
$$
$$
\theta(z_1+z_2+u_2+v_1)\theta(z_1-z_2+u_2 -v_1)
\mathcal{S}_2(z_1-1,z_2)\!=\!\theta(z_1+z_2+ u_2+v_1)\theta
(z_1-z_2+v_1-u_2)\mathcal{S}_2 (z_1,z_2-1),
$$
$$
\theta(z_1+z_2 -u_2-v_1) \theta(z_1-z_2 -u_2 +v_1)
\mathcal{S}_2 (z_1 + 1,z_2)\!=\!\theta (z_1+z_2 -u_2-v_1)
\theta(z_1-z_2 -v_1 +u_2)\mathcal{S}_2 (z_1,z_2+1),
$$
where $\theta(x)\equiv \theta_1(\eta x|\tau)$.  This leads us to
look for the solution in the form
$$\mathcal{S}_2(z_1, z_2) = \Phi_+ (z_1+z_2) \cdot \Phi_- (z_1-z_2). $$
In each equation one of the factors drops out and we obtain
coinciding difference equations for these factors. The result for
the permutation operator $\mathcal{S}_2$ in the elliptic case is
\be \label{S2e} \mathcal{S}_2 = e^{-2 \pi i \eta (u_2 - v_1) z_1}
\frac{\gamma(z_1+z_2 + u_2  - v_1 + 1)}{\gamma(z_1+z_2 - u_2+ v_1
+ 1)} \frac{\gamma(z_1-z_2 + u_2 - v_1 +1)}{\gamma(z_1-z_2 - u_2 +
v_1  + 1) }, \ee where the function $\gamma(x)$ is closely related
to the elliptic gamma-function and is defined in
Appendix~(\ref{gamma}).


We emphasize the common features allowing to do analogous steps in
the three cases for deriving the uniform expressions of operator
of parameter permutation $\mathcal{S}_2$.

\begin{prop}
The permutation operator is defined by the relation with two Lax matrices
$$
\mathcal{S}_2(u_1,u_2;v_1,v_2)\ \mathrm{L}_1(u_1,u_2)\
\mathrm{L}_2(v_1,v_2) = \mathrm{L}_1 (u_1,v_1)\
\mathrm{L}_2(u_2,v_2)\ \mathcal{S}_2(u_1,u_2;v_1,v_2),
$$
where $\mathrm{L}_i$ involves the generators of representation
with spin $\ell_i$ and is expressed in terms of the Heisenberg
pair $z_i, \dd_i$ and
$ u_1 = \frac{u}{2\eta_c} + \ell_1\ ,\ u_2 = \frac{u}{2\eta_c} -1 -\ell_1\ ;\
 v_1 = \frac{v}{2\eta_c} + \ell_2\ ,\  v_2 = 
\frac{v}{2\eta_c} -1 -\ell_2 $,
with $2\eta_0 = 2\eta_q = 1, 2\eta_e = 2\eta $. $\mathcal{S}_2$ is
represented as an operator of multiplication, depending on $z_1,
z_2$ and on $u_2-v_1$ as given in (\ref{S2o}, \ref{S2q},
\ref{S2e}).
\end{prop}
Notice that the result for the undeformed case is given by a
particular symmetric two-point function or conformal propagator.
The results in the other cases are related to the corresponding
deformed propagators.

\subsection{The operators $\mathcal{S}_1,\mathcal{S}_3$ and the
intertwining operator}

Let us consider the defining equations~(\ref{RLL13}) for the
operators $\mathcal{S}_1$ and $\mathcal{S}_3$. The permutation of
parameters $ u_1 = \frac{u}{2\eta_c} + \ell_1$ and $u_2 =
\frac{u}{2\eta_c} -1 -\ell_1 $ is equivalent to the change of the
spin $\ell_1 \to -1-\ell_1$ and similarly the permutation of
parameters $ v_1 = \frac{v}{2\eta_c} + \ell_2$ and $v_2 =
\frac{u}{2\eta_c} -1 -\ell_2$ is equivalent to the change of the
spin $\ell_2 \to -1-\ell_2$. In the Lax matrix~(\ref{Lax}) only
the generators $\mathbf{S}^a$ depend on the spin so that the defining
equations~(\ref{RLL13}) can be rewritten in terms of the
generators only. The meaning of this equations is the following:
The operator $\mathcal{S}_1$ intertwines the representations of
spin $\ell_1$ and of spin $-1-\ell_1$ and the operator
$\mathcal{S}_3$ intertwines the representations of spin $\ell_2$
and of spin $-1-\ell_2$
$$
\mathcal{S}_1\cdot \ \mathbf{S}_{\ell_1}^{a} =
\mathbf{S}_{-1-\ell_1}^{a} \ \cdot \mathcal{S}_1\ ;\
\mathcal{S}_3\cdot \ \mathbf{S}_{\ell_2}^{a} =
\mathbf{S}_{-1-\ell_2}^{a} \ \cdot \mathcal{S}_3.
$$
Let us consider the intertwining operator   $W$ for some
$\ell$, then $\mathcal{S}_1$ and  $\mathcal{S}_3$ are
special cases thereof.
\be\label{inter} W \cdot \ \mathbf{S}_{\ell}^{a} =
\mathbf{S}_{-1-\ell}^{a} \ \cdot W \ee
We study subsequently the three cases of no
deformation, quantum deformation and elliptic deformation.
\begin{itemize}
\item In the case of no deformation the equations~(\ref{inter})
can be rewritten in terms of conventional generators
$$
W \left( z \dd - \ell\right) = \left( z \dd + 1 + \ell\right) W
\quad ;\ W z \left( z \dd -2\ell \right) = z \left( z \dd
+2+2\ell\right) W\quad ;\ W\ \dd = \dd\ W
$$
The general solution to the first equation has the form $W =
z^{-2\ell-1}\cdot \Phi(z\dd)$. The two remaining equations are
compatible and lead to the difference equation:
$$
\Phi(x+1) = \frac{x+1}{x-2\ell}\cdot \Phi(x)\ ;\qquad \Phi(x) =
\frac{\Gamma(x+1)}{\Gamma(x-2\ell)}\ ,\qquad x=z\dd
$$
\item In the case of quantum deformation the
equations~(\ref{inter}) again can be rewritten in terms of
conventional generators
$$
W \left( z \dd - \ell\right) = \left( z \dd + 1 +
\ell\right) W;\ W z \left[ z \dd -2\ell \right]_q = z \left[ z
\dd +2+2\ell\right]_q W; \ W  \frac{1}{z} \left[z\dd\right]_q =
\frac{1}{z} \left[z\dd\right]_q  W
$$
The general solution of the first equation is the same as in
previous case $W = z^{-2\ell-1}\cdot \Phi(z\dd)$ and two remaining
equations result in the difference equation with the known solution
$$
\Phi(x+1)=\frac{q^{x+1}-q^{-x-1}}{q^{x-2\ell}- q^{-x+2\ell}}\cdot
\Phi(x)\ ;\ \Phi(x)=\frac{\left(q^{2x-2\ell}; q^2\right)}
{\left(q^{2x+2};q^2\right)}\cdot q^{-(2\ell+1)x}.
$$
\item In the elliptic case the intertwining operator $W$ has been
constructed  by A.Zabrodin~\cite{Zabrodin}.  For the generic
$\ell$ the solution of equations~(\ref{inter}) is given in
\cite{Zabrodin} in terms of the very well poised elliptic
hypergeometric series
\end{itemize} 

\be \label{Wel}
W = \frac{e^{2\pi i (2\ell+1) \eta z } \gamma
(2z)}{\gamma (2z+2(2\ell+1))}\cdot\sum_{k=0}^{\infty}
\frac{\left[-z-2\ell-1+2k\right]\left[-z-2\ell-1\right]_k
\left[-2\ell-1\right]_k} {\left[-z-2\ell-1\right]\left[-z+1\right]_k
\left[k\right]!} e^{(2\ell+1-2k)\dd}.
 \ee

We emphasize the common features allowing to do analogous steps in
the three cases for deriving the uniform expressions for the
intertwining operator $W$.

\begin{prop}
The intertwining operator $W(u_1,u_2)$ is defined by the relation
with Lax matrix
$$
W(u_1,u_2)\ \mathrm{L}(u_1,u_2) = \mathrm{L} (u_2,u_1)\
W(u_1,u_2),
$$
where $\mathrm{L}(u_1,u_2)$ involves the generators of
representation with spin $\ell$ and $\mathrm{L}(u_2,u_1)$ involves
the generators of representation with spin $-1-\ell$ expressed in
terms of the Heisenberg pair $z, \dd$ and
$ u_1 = \frac{u}{2\eta_c} + \ell, u_2 = \frac{u}{2\eta_c} -1 -\ell
\ ;\ u_1-u_2 = 2\ell+1 $.

In all cases the operator $W$ depends on the difference $u_1-u_2$:
$W(u_1,u_2) = W(u_1-u_2)$ \be W_o(a) =
\frac{1}{z^{a}}\cdot\frac{\Gamma(z\dd+1)}{\Gamma(z\dd+1-a)}\ ;\
W_q(a) = \frac{q^{\frac{a^2}{2}}}{z^{a}}\cdot
\frac{\left(q^{2z\dd+1-a};q^2\right)}
{\left(q^{2z\dd+2};q^2\right)} \cdot q^{-a z\dd},
\nonumber \ee
\be W_e(a) =
\frac{e^{2\pi i a \eta z+\pi i\eta a^2} \gamma (2z)}{\gamma
(2z+2a)}\cdot\sum_{k=0}^{\infty}
\frac{\left[-z-a+2k\right]\left[-z-a\right]_k \left[-a\right]_k}
{\left[-z-a\right]\left[-z+1\right]_k \left[k\right]!}
e^{(a-2k)\dd}. \nonumber \ee
\end{prop}
Note that for  future convenience we change
slightly the normalization of the operators $W_q$ and $W_e$. The
operator $\mathcal{S}_1$ is obtained from the operator
$W(u_1-u_2)$ by substitution $z\to z_1$ and the operator
$\mathcal{S}_3$ is obtained from the operator $W(v_1-v_2)$ by
substitution $z\to z_2$.

\section{Coxeter relations for the elementary 
permutation operators }
\setcounter{equation}{0}\label{proof}

In this section we shall prove that the obtained operators
$\mathcal{S}_i$ obey the Coxeter
relations~(\ref{def1}),~(\ref{def2}) and~(\ref{def3}) for the
permutation group $\mathbf{S}_4$.  The operators $\mathcal{S}_i$
have simple dependence on parameters
$$
\mathcal{S}_1(\mathbf{u}) = \mathcal{S}_1(u_1-u_2)\ ;\
\mathcal{S}_2(\mathbf{u}) = \mathcal{S}_2(u_2-v_1)\ ;\
\mathcal{S}_3(\mathbf{u}) = \mathcal{S}_3(v_1-v_2)
$$
so that the defining equations can be represented in the more
simple form \be \label{121}\mathcal{S}_i(-a)\mathcal{S}_i(a)= \II\
;\ \mathcal{S}_1(a)\mathcal{S}_2(a+b) \mathcal{S}_1(b)=
\mathcal{S}_2(b)\mathcal{S}_1(a+b) \mathcal{S}_2(a) \ee
\be\label{232} \mathcal{S}_1(a)\mathcal{S}_3(b)=
\mathcal{S}_3(b)\mathcal{S}_1(a)\ ;\
\mathcal{S}_2(a)\mathcal{S}_3(a+b) \mathcal{S}_2(b)=
\mathcal{S}_3(b)\mathcal{S}_2(a+b) \mathcal{S}_3(a). \ee There are
two evident equalities. The operator $\mathcal{S}_3(a)$ differs
from the operator $\mathcal{S}_1(a)$ only by a change of variable
$z_1\to z_2$ and therefore the operators commute,
$\mathcal{S}_1(a)\mathcal{S}_3(b)=
\mathcal{S}_3(b)\mathcal{S}_1(a).$ The equality
$\mathcal{S}_2(a)\mathcal{S}_2(-a) = \II$ can be checked easily
because the operator $\mathcal{S}_2$ reduces to
multiplication on a given function.

\subsection{The case of elliptic deformation}

The transposition operators have the form
$$
\mathcal{S}_1(a) = \frac{e^{2\pi i a \eta z_1+\pi i\eta a^2}
\gamma (2z_1)}{\gamma (2z_1+2a)}\cdot\sum_{k=0}^{\infty}
\mathrm{S}_k(a,z_1) e^{(a-2k)\dd_1}\ ;\ \mathrm{S}_k(a,z_1) =
\frac{\left[-z_1-a+2k\right]\left[-z_1-a\right]_k
\left[-a\right]_k} {\left[-z_1-a\right]\left[-z_1+1\right]_k
\left[k\right]!}$$
$$\mathcal{S}_2(a) = e^{-2\pi i a \eta z_1}
\frac{\gamma (z_1+z_2+1+a)}{\gamma(z_1+z_2+1-a)}\frac{\gamma
(z_1-z_2+1+a)}{\gamma(z_1-z_2+1-a)} , $$ where we use the
following notations for the elliptic numbers
$$[z]=\theta_1(2\eta
z)\ ;\ [z]_0=[z]\ ,\ [z]_k =[z]\cdot[z+1]\cdots[z+k-1]\ ,\
k=1,2\cdots $$ Let us calculate the product
$\mathcal{S}_1(a)\mathcal{S}_1(-a)$. Multiplying two power series
and using the formula~(\ref{transf1}) for the shifted
$\gamma$-functions we obtain the power series of the general form
with the composite coefficients
$$
\mathcal{S}_1(a)\mathcal{S}_1(-a) = \sum_{N=0}^{\infty}
\mathrm{S}_N\cdot e^{-2N\dd_1}\ ;\ \ \mathrm{S}_N = \sum_{k=0}^{N}
\mathrm{S}_k(a,z_1)\mathrm{S}_{N-k}(-a,z_1+a-2k)
\cdot\frac{\left[-z_1+1\right]_{2k}} {\left[-z_1-a+1\right]_{2k}}.
$$
We have to prove that $\mathrm{S}_0=1$ and  $\mathrm{S}_N=0$ for
$N=1,2,\cdots$. Using the formulae~(\ref{transf2}) for the
elliptic numbers we transform this expression to the canonical
form
$$
\mathrm{S}_N = \frac{\left[-z_1+2N\right] \left[-z_1\right]_N
\left[1-a-N\right]_N} {\left[-z_1\right] \left[-z_1-1+a\right]_N
\left[-N\right]_N} \cdot
$$
$$
\cdot\sum_{k=0}^{N}\frac{\left[-z_1-a+2k\right]
\left[-z_1-a\right]_k} {\left[-z_1-a\right]\left[k\right]!}\cdot
\frac{\left[-a\right]_k}{\left[-z_1+1\right]_k}
\cdot\frac{\left[-z_1+N\right]_k} {\left[1-a-N\right]_{k}}\cdot
\frac{\left[-N\right]_k}{\left[-z_1-a+1+N\right]_{k}}.
$$
The key formula which allows to calculate the sum of this special
type is the Frenkel-Turaev summation formula~\cite{FT,Rosengren}
$$
\sum_{k=0}^{N}\frac{\left[A+2k\right] \left[A\right]_k}
{\left[A\right]\left[k\right]!}\cdot
\frac{\left[B\right]_k}{\left[A+1-B\right]_k}
\cdot\frac{\left[C\right]_k}
{\left[A+1-C\right]_{k}}\cdot\frac{\left[D\right]_k}
{\left[A+1-D\right]_{k}}\cdot\frac{\left[E\right]_k}
{\left[A+1-E\right]_{k}}\cdot
\frac{\left[-N\right]_k}{\left[A+1-N\right]_{k}} =
$$
\be \label{FT} = \frac{\left[A+1\right]_N}{\left[A+1-B\right]_N}
\cdot\frac{\left[A+1-C-B\right]_N}
{\left[A+1-C\right]_{N}}\cdot\frac{\left[A+1-D-B\right]_N}
{\left[A+1-D\right]_{N}}\cdot\frac{\left[A+1-D-C\right]_N}
{\left[A+1-D-C-B\right]_{N}}. \ee
Here the coefficients are restricted by the condition
 $B+C+D+E=2A+N+1$. Our special case corresponds to the
substitutions
$$
A=-z_1-a\ ;\ B=-a\ ;\ C=-z_1+N\ ;\ D = E = \frac{A+1}{2}=
-z_1+\frac{1-a}{2}
$$
and we have
$$
\sum_{k=0}^{N}\frac{\left[-z_1-a+2k\right] \left[-z_1-a\right]_k}
{\left[-z_1-a\right]\left[k\right]!}\cdot
\frac{\left[-a\right]_k}{\left[-z_1+1\right]_k}
\cdot\frac{\left[-z_1+N\right]_k} {\left[1-a-N\right]_{k}}\cdot
\frac{\left[-N\right]_k}{\left[-z_1-a+1+N\right]_{k}} =
$$
$$
= \frac{\left[-z_1-a+1\right]_N}{\left[-z_1+1\right]_N}
\cdot\frac{\left[1-N\right]_N}
{\left[1-a-N\right]_{N}}\cdot\frac{\left[-\frac{z_1}{2}+\frac{1+a}{2}\right]_N}
{\left[-\frac{z_1}{2}+\frac{1-a}{2}\right]_{N}}\cdot
\frac{\left[\frac{z_1}{2}+\frac{1-a}{2}A-N\right]_N}
{\left[\frac{z_1}{2}+\frac{1+a}{2}A-N\right]_{N}}
$$
We obtain the needed equality $\mathrm{S}_N=0$ for $N=1,2,\cdots$
because the factor $\left[1-N\right]_N = 0$ for $N=1,2,\cdots$ and
the equality $\mathrm{S}_0=1$ can be easily checked.

Next we prove the second equality in~(\ref{121}) and the second
equality in~(\ref{232}) can be proven in a similar way.
Using~(\ref{transf1}) the second product can be transformed to the
power series of the form
$$
\mathcal{S}_2(b)\mathcal{S}_1(a+b) \mathcal{S}_2(a) =\mathrm{P}
\cdot\sum_{N=0}^{\infty} \mathrm{S}_N(a+b,z_1)\cdot
\frac{\left[-\frac{z_1+z_2}{2}+\frac{1-b}{2}\right]_N}
{\left[-\frac{z_1+z_2}{2}+\frac{1-b-2a}{2}\right]_N}
\frac{\left[-\frac{z_1-z_2}{2}+\frac{1-b}{2}\right]_N}
{\left[-\frac{z_1-z_2}{2}+\frac{1-b-2a}{2}\right]_N}
e^{(a+b-2N)\dd_1}
$$
where
$$
\mathrm{P} = e^{\pi i\eta
(b^2-a^2)}\frac{\gamma(z_1+z_2+1+2a+b)}{\gamma(z_1+z_2+1-b)}
\frac{\gamma(z_1-z_2+1+2a+b)}{\gamma(z_1-z_2+1-b)}
\frac{\gamma(2z_1)}{\gamma(2z_1+2a+2b)}.
$$
To calculate the first product we multiply two power series , use
the formula~(\ref{transf1}) and obtain the power series of the
general form
$$
\mathcal{S}_1(a)\mathcal{S}_2(a+b) \mathcal{S}_1(b)= \mathrm{P}
\cdot\sum_{N=0}^{\infty}
\left[-z_1-a-b+2N\right]\cdot\mathrm{S}_N\cdot e^{(a+b-2N)\dd_1}
$$
with the following coefficients
$$
\mathrm{S}_N = \sum_{k=0}^{N}
\mathrm{S}_k(a,z_1)\mathrm{S}_{N-k}(b,z_1+a-2k)
\cdot\frac{\left[-z_1-a-b+1\right]_{2k}}
{\left[-z_1-a+1\right]_{2k}} \cdot
\frac{\left[-\frac{z_1+z_2}{2}+\frac{1+b}{2}\right]_k}
{\left[-\frac{z_1+z_2}{2}+\frac{1-b-2a}{2}\right]_k}
\frac{\left[-\frac{z_1-z_2}{2}+\frac{1+b}{2}\right]_k}
{\left[-\frac{z_1-z_2}{2}+\frac{1-b-2a}{2}\right]_k}.
$$
Using the formulae~(\ref{transf2}) it is possible to transform
this expression to the form
$$
\mathrm{S}_N =
\frac{\left[-z_1-a-b\right]_N}{\left[-z_1-a-b\right]}
\cdot\frac{\left[1+b-N\right]_N} {\left[-N\right]_{N}}\cdot
\sum_{k=0}^{N} \frac{\left[-z_1-a+2k\right] \left[-z_1-a\right]_k}
{\left[-z_1-a\right]\left[k\right]!}\cdot
\frac{\left[-a\right]_k}{\left[-z_1+1\right]_k} \cdot
$$
$$
\cdot \frac{\left[-\frac{z_1+z_2}{2}+\frac{1+b}{2}\right]_k}
{\left[-\frac{z_1-z_2}{2}+\frac{1-b-2a}{2}\right]_k}
\frac{\left[-\frac{z_1-z_2}{2}+\frac{1+b}{2}\right]_k}
{\left[-\frac{z_1+z_2}{2}+\frac{1-b-2a}{2}\right]_k}
\frac{\left[-z_1-a-b+N\right]_k} {\left[1+b-N\right]_{k}}\cdot
\frac{\left[-N\right]_k}{\left[-z_1-a+1+N\right]_{k}}
$$
Next we use the Frenkel-Turaev formula~(\ref{FT}) for
$$
A=-z_1-a\ ;\ B=-a\ ;\ C=-\frac{z_1+z_2}{2}+\frac{1+b}{2}\ ;\
D=-\frac{z_1-z_2}{2}+\frac{1+b}{2}\ ;\ E = -z_1-a-b+N
$$
in the calculation of the sum and obtain desirable relation
$$
\mathrm{S}_N = \mathrm{S}_N(a+b,z_1)\cdot
\frac{\left[-\frac{z_1+z_2}{2}+\frac{1-b}{2}\right]_N}
{\left[-\frac{z_1+z_2}{2}+\frac{1-b-2a}{2}\right]_N}
\frac{\left[-\frac{z_1-z_2}{2}+\frac{1-b}{2}\right]_N}
{\left[-\frac{z_1-z_2}{2}+\frac{1-b-2a}{2}\right]_N}.
$$
This proves that the series expansions for the operators in both
sides of the equality~(\ref{121}) coincide.

\subsection{The case of quantum deformation}

The operators have the form \be \mathcal{S}_1(a) =
\frac{q^{\frac{a^2}{2}}}{z_1^a}\cdot
\frac{\left(q^{2-2a+2z_1\partial_1};q^2\right)}
{\left(q^{2+2z_1\partial_1};q^2\right)} \cdot q^{-a
z_1\partial_1}\ ;\ \mathcal{S}_2(a) = z_1^a
\frac{\left(\frac{z_2}{z_1}q^{1-a};q^2\right)}
{\left(\frac{z_2}{z_1}q^{1+a};q^2\right)} \ee Using the q-binomial
formula~(\ref{qbinom}) we represent the operator
$\mathcal{S}_1(a)$ as the sum \be \mathcal{S}_1(a) =
\frac{q^{\frac{a^2}{2}}}{z_1^a}\cdot\sum_{k=0}^{\infty}
\mathrm{S}_k(a) q^{-(a-2k)z_1\dd_1}\ ;\ \ \mathrm{S}_k(a) =
\frac{q^{2k}\cdot\left(q^{-2a};q^2\right)_k}{\left(q^2;q^2\right)_k}
\ee We prove the equality $\mathcal{S}_1(a)\mathcal{S}_2(a+b)
\mathcal{S}_1(b)= \mathcal{S}_2(b)\mathcal{S}_1(a+b)
\mathcal{S}_2(a)$ and the second defining equation of this type
can be proven in a similar way. Using~(\ref{transf1q}) the second
product can be transformed to the power series of the form
$$
\mathcal{S}_2(b)\mathcal{S}_1(a+b) \mathcal{S}_2(a) =
\frac{q^{\frac{b^2-a^2}{2}}\left(\frac{z_2}{z_1}q^{1-b};q^2\right)}
{\left(\frac{z_2}{z_1}q^{1+b+2a};q^2\right)}\cdot\sum_{N=0}^{\infty}
\mathrm{S}_N(a+b)\cdot
\frac{\left(\frac{z_1}{z_2}q^{1-b};q^2\right)_N}
{\left(\frac{z_1}{z_2}q^{1-b-2a};q^2\right)_N}\cdot
q^{-(a+b-2N)z_1\dd_1}
$$
To calculate the first product we multiply two power series, use the
formula~(\ref{transf1q}) and obtain the power series of the general
form
$$
\mathcal{S}_1(a)\mathcal{S}_2(a+b) \mathcal{S}_1(b)=
\frac{q^{\frac{b^2-a^2}{2}}\left(\frac{z_2}{z_1}q^{1-b};q^2\right)}
{\left(\frac{z_2}{z_1}q^{1+b+2a};q^2\right)}
\cdot\sum_{N=0}^{\infty} \mathrm{S}_N \cdot q^{-(a+b-2N)z_1\dd_1}
$$
with the following coefficients
$$
\mathrm{S}_N = q^{2N}\cdot \sum_{k=0}^{N}
\mathrm{S}_k(a)\mathrm{S}_{N-k}(b)
\cdot\frac{\left(\frac{z_1}{z_2}q^{1+b};q^2\right)_k}
{\left(\frac{z_1}{z_2}q^{1-b-2a};q^2\right)_k}\cdot q^{-2bk}.
$$
Using the formula~(\ref{transf2q}) it is possible to transform
such expression to the form
$$
\mathrm{S}_N = q^{2N}\cdot\frac{\left(q^{-2b};q^2\right)_{N}}
{\left(q^2;q^2\right)_{N}}\cdot \sum_{k=0}^{N}
\frac{\left(q^{-2a};q^2\right)_{k}}
{\left(q^2;q^2\right)_{k}}\cdot
\frac{\left(\frac{z_1}{z_2}q^{1+b};q^2\right)_k}
{\left(\frac{z_1}{z_2}q^{1-b-2a};q^2\right)_k} \cdot
\frac{\left(q^{-2N};q^2\right)_{k}}
{\left(q^{2(1+b-N)};q^2\right)_{k}}\cdot q^{2k}
$$
The key formula which allows to calculate this sum is the Jackson
summation formula~\cite{Gasper} \be \sum_{k=0}^{N}
\frac{\left(A;q\right)_{k}} {\left(q;q\right)_{k}}\cdot
\frac{\left(B;q\right)_k} {\left(C;q\right)_k} \cdot
\frac{\left(q^{-N};q\right)_{k}} {\left(\frac{A
B}{C}q^{1-N};q\right)_{k}}\cdot q^k =
\frac{\left(\frac{C}{A};q\right)_{N}} {\left(C;q\right)_{N}}\cdot
\frac{\left(\frac{C}{B};q\right)_N} {\left(\frac{C}{A
B};q\right)_N} \ee We use this formula for $q \to q^2$ and
$$
A=q^{-2a}\ ;\ B = \frac{z_1}{z_2}\cdot q^{b+1}\ ;\ C =
\frac{z_1}{z_2}\cdot q^{1-b-2a}\ ;\ \frac{A B}{C}\cdot q^{2-2N} =
q^{2+2b-2N}
$$
in the calculation of the sum and obtain the needed relation
$$
\mathrm{S}_N = q^{2N}\cdot\frac{\left(q^{-2b};q^2\right)_{N}}
{\left(q^2;q^2\right)_{N}}\cdot \frac{\left(\frac{z_1}{z_2}\cdot
q^{1-b};q^2\right)_{N}} {\left(\frac{z_1}{z_2}\cdot
q^{1-b-2a};q^2\right)_{N}}\cdot \frac{\left(q^{-a-b};q^2\right)_N}
{\left(q^{-2b};q^2\right)_N} = \mathrm{S}_N(a+b)\cdot
\frac{\left(\frac{z_1}{z_2}q^{1-b};q^2\right)_N}
{\left(\frac{z_1}{z_2}q^{1-b-2a};q^2\right)_N}.
$$This proves that the series expansions for the operators in both
sides of the equality~(\ref{121}) coincide.

\subsection{The case of no deformation}

The operators have the form ($(a)_k = a(a+1)\cdots (a+k-1)$)\be
\mathcal{S}_1(a) = \frac{1}{z_1^a}\cdot
\frac{\Gamma\left(z_1\partial_1
+1\right)}{\Gamma\left(z_1\partial_1 +1-a\right)} \ ;\
\mathcal{S}_2(a) = \left(z_2-z_1\right)^a =
z_2^a\cdot\sum_{k=0}^{\infty} \frac{\left(-a \right)_k}{k!}\cdot
\left(\frac{z_1}{z_2}\right)^k \ee We again prove the equality
$\mathcal{S}_1(a)\mathcal{S}_2(a+b) \mathcal{S}_1(b)=
\mathcal{S}_2(b)\mathcal{S}_1(a+b) \mathcal{S}_2(a)$ only. Using
the equality
\begin{equation}\label{transf10}
\Gamma(z_1\partial_1 + A )\cdot z_1^n = z_1^n \cdot
\Gamma(z_1\partial_1 + A+n )
\end{equation}
the first product can be transformed to the power series of the
form
$$
\mathcal{S}_1(a)\mathcal{S}_2(a+b) \mathcal{S}_1(b) =
\sum_{N=0}^{\infty} \frac{\left(-a -b\right)_N}{N!}\cdot
\left(\frac{z_1}{z_2}\right)^{N-a-b}
\cdot\frac{\Gamma\left(z_1\partial_1
+1-b+N\right)}{\Gamma\left(z_1\partial_1
+1-a-b+N\right)}\cdot\frac{\Gamma\left(z_1\partial_1
+1\right)}{\Gamma\left(z_1\partial_1 +1-b\right)}
$$
To calculate the second product we multiply two power series , use
the formula~(\ref{transf10}) and obtain the power series of the
general form with the composite coefficients
$$
\mathcal{S}_2(b)\mathcal{S}_1(a+b) \mathcal{S}_2(a)=
\sum_{N=0}^{\infty} \left(\frac{z_1}{z_2}\right)^{N-a-b}\cdot
\mathrm{S}_N \ ;\ \mathrm{S}_N = \sum_{k=0}^{N}
\frac{\left(-a\right)_k}{k!}\cdot
\frac{\left(-b\right)_{N-k}}{(N-k)!}
\cdot\frac{\Gamma\left(z_1\partial_1
+1+k\right)}{\Gamma\left(z_1\partial_1 +1-a-b+k\right)}
$$
Using the formula
\begin{equation}
\frac{\left(-b\right)_{N-k}} {(N-k)!} = \frac{\left(-b\right)_{N}}
{N!}\cdot \frac{\left(-N\right)_{k}} {\left(1+b-N\right)_{k}}
\end{equation}
it is possible to transform such expression to the form
($\mathbf{d}\equiv z_1\partial_1$)
$$
\mathrm{S}_N = \sum_{k=0}^{N} \frac{\left(-a\right)_k}{k!}\cdot
\frac{\left(\mathbf{d}+1\right)_{k}}{\left(\mathbf{d}+1-a-b\right)_{k}}
\cdot \frac{\left(-N\right)_{k}} {\left(1+b-N\right)_{k}}
\cdot\frac{\left(-b\right)_{N}} {N!}\cdot
\frac{\Gamma\left(\mathbf{d} +1\right)}{\Gamma\left(\mathbf{d}
+1-a-b\right)}
$$
The key formula which allows to calculate the sum of this special
type is the Pfaff-Saalschutz summation formula~\cite{Gasper} \be
\sum_{k=0}^{N} \frac{\left(A\right)_{k}} {k!}\cdot
\frac{\left(B\right)_k} {\left(C\right)_k} \cdot
\frac{\left(-N\right)_{k}} {\left(1+A+B-C-N\right)_{k}} =
\frac{\left(C-A\right)_{N}} {\left(C\right)_{N}}\cdot
\frac{\left(C-B\right)_N} {\left(C - A - B\right)_N} \ee
We use
this formula for
$$
A = -a\ ;\ B = \mathbf{d}+1\ ;\ C = \mathbf{d}+1-a-b\ ;\ 1+A+B-C
-N = 1-b-N
$$
in the calculation of the sum and obtain the needed relation
$$
\mathrm{S}_N = \frac{\left(-a
-b\right)_N}{N!}\cdot\frac{\Gamma\left(\mathbf{d}
+1-b+N\right)}{\Gamma\left(\mathbf{d}
+1-a-b+N\right)}\cdot\frac{\Gamma\left(\mathbf{d}
+1\right)}{\Gamma\left(\mathbf{d} +1-b\right)}.
$$ This proves that the series expansions for the operators in both
sides of the equality~(\ref{121}) coincide.

\section{The permutation group and the Yang-Baxter relation}
\setcounter{equation}{0}

Let us return to the permutation group. The transpositions
$s_1,s_2,s_3$ form the basis of the permutation group of four
parameters $(u_1,u_2,v_1,v_2)$. Any permutation can be decomposed
into the product of these operators $s_k$. This representation is
not unique because different decompositions can represent the
same permutation. The equivalence of the permutation group
$\mathfrak{S}_4$ and the Coxeter group with the defining relations
$s^2_1 = s^2_2=\II$ , $s_1s_3 = s_3s_1$ and $s_1s_2s_1 =
s_2s_2s_2$, $s_2s_3s_2 =s_3s_2s_3$ guarantees that two sequences
representing the same permutation can be transformed each other by
using the defining relations between generators only. In the
general case it is the existence theorem, but we shall
follow the more
constructive way and indicate the particular
transformations we need each time.
 The operators
$\mathcal{S}_1,\mathcal{S}_{2},\mathcal{S}_{3}$ represent the 
corresponding
permutations of four parameters $(u_1,u_2,v_1,v_2)$ entering in
the product of two Lax matrices
$\mathrm{L}_1(u_1,u_2)\mathrm{L}_2(v_1,v_2)$ and obey the same
defining relations. This key property of the operators 
$\mathcal{S}_i$
shows that we deal with a representation of the permutation group.

Note that everything can be generalized to the product of an
arbitrary number of Lax matrices. We consider the product of the
three Lax matrices
$\mathrm{L}_1(u_1,u_2)\mathrm{L}_2(v_1,v_2)\mathrm{L}_3(w_1,w_2)$
because this example immediately leads to the Yang-Baxter
equation. Let us join all six parameters in one set
$\mathbf{u}\equiv(u_1,u_2,v_1,v_2,w_1,w_2)$ and consider the group
of permutations $\mathbf{S}_6$ of six parameters. Now we have five
elementary transpositions $s_i$ and corresponding operators
$\mathcal{S}_i(\mathbf{u})$
$$
(\overset{\mathcal{S}_1}{\overbrace{u_1\ ,\ u_2}},
\overset{\mathcal{S}_{3}}{\overbrace{v_1\ ,\
v_2}},\overset{\mathcal{S}_{5}}{\overbrace{w_1\ ,\ w_2}})\ ;\ (u_1
, \overset{\mathcal{S}_2}{\overbrace{u_2\ ,\ v_1}} ,
\overset{\mathcal{S}_{4}}{\overbrace{v_2\ ,\ w_1}} , w_2),
$$
which obey the following defining relations
$$
\mathcal{S}_1\,\mathrm{L}_1(u_1,u_2) =
\mathrm{L}_1(u_2,u_1)\mathcal{S}_1\ ;\
\mathcal{S}_3\,\mathrm{L}_2(v_1,v_2) =
\mathrm{L}_2(v_2,v_1)\mathcal{S}_3\ ;\
\mathcal{S}_5\,\mathrm{L}_3(w_1,w_2) =
\mathrm{L}_2(w_2,w_1)\mathcal{S}_5
$$
$$
\mathcal{S}_2(\mathbf{u})\,\mathrm{L}_1(u_1,u_2)\,\mathrm{L}_2(v_1,v_2)=
\mathrm{L}_1(u_1,v_1)\,\mathrm{L}_2(u_2,v_2)\mathcal{S}_2(\mathbf{u})\,,
$$
$$
\mathcal{S}_4(\mathbf{u})\,\mathrm{L}_2(v_1,v_2)\,\mathrm{L}_3(w_1,w_2)=
\mathrm{L}_2(v_1,w_1)\,\mathrm{L}_3(v_2,w_2)\mathcal{S}_4(\mathbf{u})\,.
$$
It is evident that all is effectively reduced to the case of the
product of two Lax matrices. The generators $\mathcal{S}_1$ ,
$\mathcal{S}_2$ and $\mathcal{S}_3$ are the same as in the case of
the product of two Lax matrices
$\mathrm{L}_1(u_1,u_2)\mathrm{L}_2(v_1,v_2)$. The generators
$\mathcal{S}_3$ , $\mathcal{S}_4$ and $\mathcal{S}_5$ play the
same role for the product
$\mathrm{L}_2(v_1,v_2)\mathrm{L}_3(w_1,w_2)$ and can be obtained
from the $\mathcal{S}_1$ , $\mathcal{S}_2$ and $\mathcal{S}_3$ by
the simple change of variables and parameters
\begin{equation}\label{change}
z_1,z_2 \rightarrow z_2,z_3\ ;\ (u_1,u_2,v_1,v_2)\rightarrow
(v_1,v_2,w_1,w_2).
\end{equation}
The defining relations
$$
s_i s_i = 1 \rightarrow
\mathcal{S}_i(s_i\mathbf{u})\mathcal{S}_i(\mathbf{u})= 1\ ;\
s_is_j = s_js_i \rightarrow
\mathcal{S}_i(s_j\mathbf{u})\mathcal{S}_j(\mathbf{u})=
\mathcal{S}_j(s_i\mathbf{u})\mathcal{S}_i(\mathbf{u})\ , |i-j|>1
$$
$$
s_i s_{i+1} s_i = s_{i+1} s_i s_{i+1} \rightarrow
\mathcal{S}_i(s_{i+1}s_i\mathbf{u})\mathcal{S}_{i+1}(s_i\mathbf{u})
\mathcal{S}_i(\mathbf{u})=
\mathcal{S}_{i+1}(s_is_{i+1}\mathbf{u})\mathcal{S}_i(s_{i+1}\mathbf{u})
\mathcal{S}_{i+1}(\mathbf{u})
$$ are effectively reduced to the
relations for the operators from the previous case also. The
operator $\check{\mathcal{R}}_{12}(u-v)$ is the solution of the
equation
$$
\check{\mathcal{R}}_{12}(u-v)\,\mathrm{L}_1(u_1,u_2)\,\mathrm{L}_2(v_1,v_2)=
\mathrm{L}_1(v_1,v_2)\,\mathrm{L}_2(u_1,u_2)\,\check{\mathcal{R}}_{12}(u-v)
$$
and corresponds to the permutation $s_2s_1s_3s_2$:
\be \label{R12}
\check{\mathcal{R}}_{12}(u-v)= \mathcal{S}_2(s_1s_3s_2\mathbf{u})
\mathcal{S}_1(s_3s_2\mathbf{u})\mathcal{S}_3(s_2\mathbf{u})
\mathcal{S}_2(\mathbf{u}).
\ee
The operator
$\check{\mathcal{R}}_{23}(v-w)$ is the solution of the equation
$$
\check{\mathcal{R}}_{23}(v-w)\,\mathrm{L}_2(v_1,v_2)\,\mathrm{L}_3(w_1,w_2)=
\mathrm{L}_2(w_1,w_2)\,\mathrm{L}_3(v_1,v_2)\,\check{\mathcal{R}}_{23}(v-w)
$$
and corresponds to the permutation $s_4s_3s_5s_4$:
\be \label{R23}
\check{\mathcal{R}}_{23}(v-w)= \mathcal{S}_4(s_3s_5s_4\mathbf{u})
\mathcal{S}_3(s_5s_4\mathbf{u})\mathcal{S}_5(s_4\mathbf{u})
\mathcal{S}_4(\mathbf{u}).
\ee
 The operator
$\check{\mathcal{R}}_{23}(v-w)$ can be obtained by the same change
of variables and parameters~(\ref{change}) from the operator
$\check{\mathcal{R}}_{12}(u-v)$. There exists two equivalent
decompositions of the special permutation
$$
s(u_1,u_2,v_1,v_2,w_1,w_2)=(w_1,w_2,v_1,v_2,u_1,u_2)
$$
in terms of permutations $s_2s_1s_3s_2$ and $s_4s_3s_5s_4$ and
corresponding operators are expressed in terms of R-matrix
$$
s= s_2s_1s_3s_2\cdot s_4s_3s_5s_4\cdot s_2s_1s_3s_2 \rightarrow
\check{\mathcal{R}}_{12}(v-w)
\check{\mathcal{R}}_{23}(u-w)\check{\mathcal{R}}_{12}(u-v),
$$
$$
s= s_4s_3s_5s_4\cdot s_2s_1s_3s_2 \cdot s_4s_3s_5s_4 \rightarrow
\check{\mathcal{R}}_{23}(u-v)
\check{\mathcal{R}}_{12}(u-w)\check{\mathcal{R}}_{23}(v-w).
$$
These operators correspond the same permutation and therefore
there exists the transformation from the one expression to the
another.
\begin{prop}
The operators $\check{\mathcal{R}}_{12}$ (\ref{R12}) and
$\check{\mathcal{R}}_{23}$ (\ref{R23})
obey the Yang-Baxter relation
$$
\check{\mathcal{R}}_{23}(u-v)
\check{\mathcal{R}}_{12}(u-w)\check{\mathcal{R}}_{23}(v-w) =
\check{\mathcal{R}}_{12}(v-w)
\check{\mathcal{R}}_{23}(u-w)\check{\mathcal{R}}_{12}(u-v).
$$
\end{prop}
It is sufficient to give an example of the chain 
of transformations allowing to
transform the first decomposition of the permutation $s$ to the
second one
$$
s_2s_3\underline{s_1s_2}\cdot \underline{s_4}s_3s_5\underline{s_4}
\cdot \underline{s_2s_1}s_3s_2 = s_2s_3s_4\cdot
s_1\underline{s_2s_3s_2}s_1\cdot s_5 \cdot s_4s_3s_2 =
s_2\underline{s_3s_4s_3}\cdot\underline{s_1s_2s_1}\cdot s_5 \cdot
\underline{s_3s_4s_3}s_2 =
$$
$$
\underline{s_2s_4}s_3\underline{s_4}\cdot
\underline{s_2}s_1\underline{s_2}\cdot s_5 \cdot
\underline{s_4}s_3\underline{s_4s_2} =
s_4\underline{s_2s_3s_2}\cdot s_1\cdot \underline{s_4s_5s_4} \cdot
\underline{s_2s_3s_2}s_4 = s_4s_3\underline{s_2s_3}\cdot s_1\cdot
\underline{s_5}s_4\underline{s_5} \cdot \underline{s_3s_2}s_3s_4 =
$$
$$
s_4s_3s_5\cdot s_2s_1 \underline{s_3s_4 s_3}s_2\cdot s_5s_3s_4
=s_4s_3s_5\cdot s_2s_1 \underline{s_4}s_3 \underline{s_4}s_2\cdot
s_5s_3s_4 = s_4s_3s_5s_4\cdot s_2s_1s_3s_2\cdot s_4s_5s_3s_4
$$
Repeating step by step this chain of transformations for the
operators $\mathcal{S}_i$ it is possible to transform the operator
in the right hand side of Yang-Baxter relation to the operator in
the left hand side. We omit the corresponding formulae for 
brevity.

\section{The operators $\check{\mathcal{R}}^{(1)}$ and
$\check{\mathcal{R}}^{(2)}$}
\setcounter{equation}{0}

Recall the different levels of resolution of the Yang-Baxter
operator into operators representing more elementary 
parameter permutations in acting on the product of Lax matrices
as discussed in Introduction. Besides of the operators 
$\mathcal{S}_i$ representing the elementary transpositions the following
operators $\check{\mathcal{R}}^{(k)}, k= 1,2$ are important; they
permute the parameters $u_k, v_k$ in acting on the product of
Lax matrices. 

Again everything can be generalized to the product of arbitrary
number of Lax matrices. We consider the product of the three Lax
matrices
$\mathrm{L}_1(u_1,u_2)\mathrm{L}_2(v_1,v_2)\mathrm{L}_3(w_1,w_2)$
because this example immediately leads to the analogon of the
Yang-Baxter relations for the operators
$\check{\mathcal{R}}^{(i)}$, $i=1,2$. Recall that working in terms
of operators $\check{\mathcal{R}}^{(i)}$ we effectively reduce the
symmetry $\mathfrak{S}_6 \rightarrow
\mathfrak{S}_3\times\mathfrak{S}_3$.

Let us introduce the special permutations $r_1 = s_2 s_{1} s_2$,
$r_3 = s_4 s_{3} s_4$ and $p_3 = s_2 s_{3} s_2$, $p_5 = s_4 s_{5}
s_4$. The permutation $r_1$ and $r_3$ interchange only the
parameters $u_1\leftrightarrow v_1$ and $v_1\leftrightarrow w_1$
correspondingly
$$
r_1\mathbf{u} = s_2 s_{1} s_2\mathbf{u} =
(v_1,u_2,u_1,v_2,w_1,w_2)\ ;\ r_3\mathbf{u} = s_4 s_{3}
s_4\mathbf{u} = (u_1,u_2,w_1,v_2,v_1,w_2)
$$
and generate the group of permutations $\mathfrak{S}_3$ of three
parameters~$(u_1,v_1,w_1)$. The corresponding operators have the
form
\be \label{R1}
 \check{\mathcal{R}}^{(1)}_{12}(\mathbf{u})=
\mathcal{S}_2(s_1s_2\mathbf{u})\mathcal{S}_1(s_2\mathbf{u})
\mathcal{S}_2(\mathbf{u})\ ;\
\check{\mathcal{R}}^{(1)}_{23}(\mathbf{u})=
\mathcal{S}_4(s_3s_4\mathbf{u})\mathcal{S}_3(s_4\mathbf{u})
\mathcal{S}_4(\mathbf{u})
\ee
The permutation $p_3$ and $p_5$ interchange in a similar way
$u_2\leftrightarrow v_2$ and $v_2\leftrightarrow w_2$
$$
p_3\mathbf{u} = s_2 s_{3} s_2\mathbf{u}
=(u_1,v_2,v_1,u_2,w_1,w_2)\ ;\ p_5\mathbf{u} = s_4 s_{5}
s_4\mathbf{u} = (u_1,u_2,v_1,w_2,w_1,v_2)
$$
and generate the group of permutations $\mathfrak{S}_3$ of three
parameters~$(u_2,v_2,w_2)$. The corresponding operators are
\be \label{R2}
\check{\mathcal{R}}^{(2)}_{12}(\mathbf{u})=
\mathcal{S}_2(s_3s_2\mathbf{u})\mathcal{S}_3(s_2\mathbf{u})\mathcal{S}_2(\mathbf{u})
\ ;\ \check{\mathcal{R}}^{(2)}_{23}(\mathbf{u})=
\mathcal{S}_4(s_5s_4\mathbf{u})\mathcal{S}_5(s_4\mathbf{u})\mathcal{S}_4(\mathbf{u})
\ee

\begin{prop}
There are the following equivalent representations for the
operator $\check{\mathcal{R}}_{12}$
$$
\check{\mathcal{R}}_{12}(u-v) =
\check{\mathcal{R}}^{(1)}_{12}(p\mathbf{u})\check{\mathcal{R}}^{(2)}_{12}(\mathbf{u})=
\check{\mathcal{R}}^{(2)}_{12}(r\mathbf{u})\check{\mathcal{R}}^{(1)}_{12}(\mathbf{u})=
\mathcal{S}_2(s_1s_3s_2\mathbf{u})
\mathcal{S}_1(s_3s_2\mathbf{u})\mathcal{S}_3(s_2\mathbf{u})
\mathcal{S}_2(\mathbf{u}).
$$
\end{prop}

The first and second expressions correspond to the decompositions
of the permutation $s$ on the product of two commuting
permutations $r=s_2 s_1 s_2$ and $p=s_2 s_3 s_2$. The third
expression corresponds to the decomposition $s = s_2 s_1s_3s_2$.
The chain of the transformations is
$$
r p = s_2s_1s_2\cdot s_2s_3s_2 = s_2s_1\underline{s_2\cdot
s_2}s_3s_2 = s_2s_1s_3s_2 = s_2s_3s_1s_2 =
s_2s_1\underline{s_2\cdot s_2}s_3s_2 = p r
$$
and there is the corresponding chain of transformations which
proves the equivalence of all representations for the R-matrix
$$
\check{\mathcal{R}}^{(1)}_{12}(p\mathbf{u})\check{\mathcal{R}}^{(2)}_{12}(\mathbf{u})
= \mathcal{S}_2(s_1s_2\cdot s_2
s_3s_2\mathbf{u})\mathcal{S}_1(s_2\cdot s_2 s_3s_2\mathbf{u})
\underline{\mathcal{S}_2(s_2 s_3s_2\mathbf{u})\cdot
\mathcal{S}_2(s_3s_2\mathbf{u})}\mathcal{S}_3(s_2\mathbf{u})\mathcal{S}_2(\mathbf{u})=
$$$$
=\mathcal{S}_2(s_1s_3s_2\mathbf{u})\underline{\mathcal{S}_1(s_3s_2\mathbf{u})
\mathcal{S}_3(s_2\mathbf{u})}\mathcal{S}_2(\mathbf{u})=
\mathcal{S}_2(s_3s_1s_2\mathbf{u})\mathcal{S}_3(s_1s_2\mathbf{u})
\mathcal{S}_1(s_2\mathbf{u})\mathcal{S}_2(\mathbf{u}) =
$$
$$
=\mathcal{S}_2(s_3s_2\cdot
s_2s_1s_2\mathbf{u})\mathcal{S}_3(s_2\cdot s_2s_1s_2\mathbf{u})
\underline{\mathcal{S}_2(s_2s_1s_2\mathbf{u})\cdot\mathcal{S}_2(s_1s_2\mathbf{u})}
\mathcal{S}_1(s_2\mathbf{u})\mathcal{S}_2(\mathbf{u}) =
\check{\mathcal{R}}^{(2)}_{12}(r\mathbf{u})
\check{\mathcal{R}}^{(1)}_{12}(\mathbf{u}).
$$

\begin{prop}
The operators $\check{\mathcal{R}}^{(1)}$ (\ref{R1}) and
$\check{\mathcal{R}}^{(2)}$ (\ref{R2}) obey the relations
$$
\check{\mathcal{R}}^{(2)}_{12}(r_1 \mathbf{u})
\check{\mathcal{R}}^{(1)}_{12}(\mathbf{u}) =
\check{\mathcal{R}}^{(1)}_{12}(p_3\mathbf{u})
\check{\mathcal{R}}^{(2)}_{12}(\mathbf{u})\ ;\
\check{\mathcal{R}}^{(2)}_{23}(r_3 \mathbf{u})
\check{\mathcal{R}}^{(1)}_{23}(\mathbf{u}) =
\check{\mathcal{R}}^{(1)}_{23}(p_5\mathbf{u})
\check{\mathcal{R}}^{(2)}_{23}(\mathbf{u})
$$
$$
\check{\mathcal{R}}^{(2)}_{23}(r_1 \mathbf{u})
\check{\mathcal{R}}^{(1)}_{12}(\mathbf{u}) =
\check{\mathcal{R}}^{(1)}_{12}(p_5\mathbf{u})
\check{\mathcal{R}}^{(2)}_{23}(\mathbf{u})\ ;\
\check{\mathcal{R}}^{(2)}_{12}(r_3 \mathbf{u})
\check{\mathcal{R}}^{(1)}_{23}(\mathbf{u}) =
\check{\mathcal{R}}^{(1)}_{23}(p_3\mathbf{u})
\check{\mathcal{R}}^{(2)}_{12}(\mathbf{u})
$$
$$
\check{\mathcal{R}}^{(1)}_{23}(r_1r_3\mathbf{u}) \check{\mathcal{R}}
^{(1)}_{12}(r_3\mathbf{u})\check{\mathcal{R}}^{(1)}_{23}(\mathbf{u})
=\check{\mathcal{R}}^{(1)}_{12}(r_3r_1\mathbf{u})\check{\mathcal{R}}
^{(1)}_{23}(r_1\mathbf{u})\check{\mathcal{R}}^{(1)}_{12}(\mathbf{u}).
$$
$$
\check{\mathcal{R}}^{(2)}_{23}(p_3p_5\mathbf{u}) \check{\mathcal{R}}
^{(2)}_{12}(p_5\mathbf{u})\check{\mathcal{R}}^{(2)}_{23}(\mathbf{u})
=\check{\mathcal{R}}^{(2)}_{12}(p_5p_3\mathbf{u})\check{\mathcal{R}}
^{(2)}_{23}(p_3\mathbf{u})\check{\mathcal{R}}^{(2)}_{12}(\mathbf{u}).
$$
\end{prop}
These relations correspond to the relations for the group of
permutations $\mathfrak{S}_3\times\mathfrak{S}_3$. The first four are
relations of commutativity for the transformations from two
different groups $\mathfrak{S}_3$
$$
p_3 r_1 = r_1 p_3\ ;\ p_5 r_3 = r_3 p_5\ ;\ p_5 r_1 = r_1 p_5\ ;\
p_3 r_3 = r_3 p_3
$$
The proof of the corresponding relations for the operators
$\check{\mathcal{R}}^{(i)}$ mimics step by step the
transformations for the permutation group. We collect these chains
of transformations in the same order as they are listed above.
$$
p_3 r_1 = s_2s_3s_2\cdot s_2s_1s_2 = s_2s_3s_1s_2 = s_2s_1s_3s_2 =
s_2s_1s_2\cdot s_2s_3s_2 = r_1 p_3
$$
$$
p_5 r_3 = s_4s_5s_4\cdot s_4s_3s_4 = s_4s_5s_3s_4 = s_4s_3s_5s_4 =
s_4s_3s_4\cdot s_4s_5s_4 = r_3 p_5
$$
$$
p_5 r_1 = s_4s_5s_4\cdot s_2s_1s_2 = s_2s_1s_2\cdot s_4s_5s_4= r_1
p_5
$$
$$
p_3 r_3 = s_2s_3s_2\cdot s_4s_3s_4 = s_3s_2s_3\cdot s_3s_4s_3 =
s_3s_2s_4s_3 = s_3s_4s_2s_3 = s_3s_4s_3\cdot s_3s_2s_3 =
s_4s_3s_4\cdot s_2s_3s_2 = r_3 p_3
$$
The last two relations are the triple defining relations
of each group $\mathfrak{S}_3$
$$
r_3 r_1 r_3 = r_1 r_3 r_1\ ;\ p_5 p_3 p_5 = p_3 p_5 p_3
$$
and the chains of transformations have the form
$$
r_1 r_3 r_1 = s_2s_1s_2\cdot s_4s_3s_4\cdot s_2s_1s_2 =
s_1s_2s_1\cdot s_4s_3s_4\cdot s_1s_2s_1 = s_4 s_1 \cdot
s_2s_3s_2\cdot s_1s_4 =  s_4 s_1\cdot s_3s_2s_3 \cdot s_1s_4 =
$$
$$
= s_4s_3\cdot s_1s_2s_1\cdot s_3s_4 = s_4s_3s_4\cdot
s_2s_1s_2\cdot s_4s_3s_4 = r_3 r_1 r_3
$$
$$
p_3 p_5 p_3 = s_2s_3s_2\cdot s_4s_5s_4\cdot s_2s_3s_2 =
s_2s_3s_2\cdot s_5s_4s_5\cdot s_2s_3s_2 = s_5s_2\cdot s_3 s_4
s_3\cdot s_2s_5 = s_5s_2\cdot s_4 s_3 s_4\cdot s_2s_5 =
$$
$$
= s_5s_4s_5\cdot s_2s_3s_2\cdot s_5s_4s_5 = s_4s_5s_4\cdot
s_2s_3s_2\cdot s_4s_5s_4 = r_5 r_3 r_5
$$

\section{Discussion and summary}
\setcounter{equation}{0}

In the present analysis we did not specify to a series of group
representations related to the considered Lie algebra $s\ell_2 $. 
This would imply a restriction 
on the values of the representation parameter $\ell$, but this appears
unnatural here because $\ell$ enters the relevant expressions in 
combination with the spectral parameter. 

In the undeformed and q-deformed cases the representation $\ell$ 
can be realized as a module spanned by the monomials $z^m, m= 0,1,...$,
not assuming a metric structure of a functional space. The constsnt f
undtion$1$
represents the lowest weight vector. An invariant finite-dimensional
sub-modul appears for positive integer values of $2\ell$. 
In the elliptic case the representation can be described in terms of
entire even periodic functions  and the finite-dimensional
representation appearing in the case of positive integer $N=2\ell$ is 
spanned by even products of $2N$ $\theta $-functions.
These realizations are convenient for describing the generic 
representations and their tensor products. 
The Yang-Baxter operator and its factors
$\mathcal{R}^{(1)}, \mathcal{R}^{(2)} $ are operating in these classes       
of functions. However the operators of elementary permutations $
\mathcal{S}_i$ map to different  realizations involving e.g. functions 
with branch points.

Also the ambiguity of the solutions of difference equations by 
periodic factors is to be fixed by specifying the function class.

We refer to the
particular case of the principal series of $SL(2, \C) $ representations
where all operators in question can be defined as acting
on functions on the complex plane \cite{DKM} and the mentioned
difficulty does not appear.   
There one deals with actually two copies of $s\ell_2$, one represented by
operators acting holomorphically and the other anti-holomorphically on 
functions defined on the complex plane. 
Correspondingly, the representations are
labelled by the pair $(\ell, \bar \ell)$ taking values
$\ell = \frac{n-1}{2} - i\nu, \bar \ell = - \frac{n+1}{2} -i \nu $,
$n$ integer and $\nu$ real.  The structure of the operators being
products of a holomorphic and anti-holomorphic part ensures that
they do not lead beyond uniquely defined functions on the complex plane.

The simplest case of a symmetry algebra of the Yang-Baxter equation
is $s\ell(2)$ and there are two possible ways of generalization.
First one can  deform the algebraic structure but keep the rank
equal to one:
$$
U(s\ell_2) \rightarrow U_q(s\ell_2) \rightarrow {\rm Sklyanin \
algebra}
$$
In this paper we present the uniform expression for the solution
of the Yang-Baxter equation connected with all three cases. The
R-operator acting in the tensor product of two representations of
the symmetry algebra with spins $\ell_1$ and $\ell_2$ can be
constructed from the three basic operators $\mathcal{S}_1,
\mathcal{S}_2,\mathcal{S}_3$. The operators $\mathcal{S}_1,
\mathcal{S}_2,\mathcal{S}_3$ represent elementary permutations
from the symmetric group $\mathfrak{S}_4$ -- permutation group of
four parameters entering the RLL-relation.

The second way of generalization is to increase the rank going to
the algebra $s\ell(n)$. The Weyl group $W$ of $s\ell_n$ is the
permutation group $\mathfrak{S}_n$ generated by elements
$w_1,\ldots, w_{n-1}$ with defining relations $w_i^2=1$,
$w_iw_{i+1}w_i=w_{i+1}w_iw_{i+1}$ and $w_iw_j=w_jw_i$ if $|i-j|>1$.
The RLL-relation has the same form~(\ref{RLL})
$$
\check{\mathcal{R}}_{12}(u-v)\,\mathrm{L}_1(u_1,u_2\cdots
u_n)\,\mathrm{L}_2(v_1,v_2\cdots v_n)= \mathrm{L}_1(v_1,v_2\cdots
v_n)\,\mathrm{L}_2(u_1,u_2\cdots
u_n)\,\check{\mathcal{R}}_{12}(u-v)\ ,
$$
but now the Lax matrix depends on the n parameters $u_1,u_2\cdots
u_n$~\cite{DM}. There are $n-1$ parameters which label the
representation of the algebra $s\ell(n)$ (analog of spin $\ell$)
and spectral parameter $u$. There are $2n$ parameters in the
RLL-relation and the permutation group  is now $\mathfrak{S}_{2n}$.
There are $2n-1$ elementary transpositions in the group
$\mathfrak{S}_{2n}$
$$
\overset{s_1}{\overbrace{u_1,u_2}}\cdots
\overset{s_{n-1}}{\overbrace{u_{n-1} , u_n}}
,\overset{s_{n+1}}{\overbrace{v_1, v_2}}\cdots
\overset{s_{2n-1}}{\overbrace{v_{n-1} , v_n}} \ \ ;\ \
u_1,u_2\cdots u_{n-1} , \overset{s_{n}}{\overbrace{u_n , v_1}}
,v_2\cdots v_{n-1} , v_n
$$
The R-matrix again admits the decomposition in factors of operators
representing the elementary transpositions~\cite{DM}. The
operators, representing $s_1,\cdots,s_{n-1}$ and
$s_{n+1},\cdots,s_{2n-1}$ are the well known intertwining operators
and are connected to the Weyl group of the algebra $s\ell(n)$. The
operator representing $s_n$ plays a special role and can be
constructed explicitly~\cite{DM}.

It seems that the considered  decomposition of the R-matrix is
universal and gives some insight into a possible structure of
solutions of Yang-Baxter equation with general symmetry.

In terms of the elementary parameter permutation operators
we construct the factor operators  $\mathcal{R}^{(k)}$.
The 
product of these  factors over $k=1, ..,n$ results 
in the Yang-Baxter  $\RR$ operator for $s\ell(n)$. 
Acting on the product of two 
Lax operators $\mathcal{R}^{(k)}$ permutes their parameters
$u_k, v_k$. For a spin chain the products of these factor 
operators over the sites 
leads to a commuting family of operators $\mathcal{Q}_k$ 
which can be identified as the Baxter operators~\cite{D,DM1,DKK}.

\section*{Acknowledgement}

We thank A. Zabrodin for  illuminating discussions and
explanations.

This work has been supported by the RFFI grant 05-01-00922,
partially by  grant NSh-5403.2006.1
and by DFG grant 436 Rus 17/9/06 (S.D.), 
by DFG grant 436 Arm 17/1/06, by Volkswagen Stiftung (D.K.) and 
by NTZ of Leipzig University.

\section{Appendix}
\setcounter{equation}{0}

\subsection{q-special functions}

The standard q-products are
$$
(x;q) = \prod_{k=0}^{+\infty}(1-q^k\cdot x) \ ;\ (x;q)_{n} =
\prod_{k=0}^{n-1}(1-q^k\cdot x) = \frac{(x;q)}{(xq^n;q)} \ ;\ q\in
\C \ ,\ |q| < 1
$$
The q-binomial formula has the form \be \label{qbinom}
\frac{\left(Az;q\right)}{\left(z;q\right)} = \sum_{k=0}^{\infty}
\frac{\left(A;q\right)_k}{\left(q;q\right)_k}\cdot z^k \ee The
function $(x;q)$ obeys the recurrent relation $ (qx;q)
=(1-x)^{-1}(x;q)$ so that it is used to solve the recurrent
relation of the type \be\label{recq} \Phi(q x) = \frac{1-ax}{1-bx}
\Phi(x) \ ;\ \Phi(x) = \frac{(bx;q)}{(ax;q)}\ee There are useful
formulae for the q-products which are used in the text
\begin{equation}\label{transf1q}
\frac{\left(yq^{-2n};q^2\right)}{\left(xq^{-2n};q^2\right)} =
\frac{y^n}{x^n}\cdot
\frac{\left(\frac{q^2}{y};q^2\right)_n}{\left(\frac{q^2}{x};q^2\right)_n}
\frac{\left(y;q^2\right)}{\left(x;q^2\right)}
\end{equation}
\begin{equation}\label{transf2q}
\frac{\left(q^{-2b};q^2\right)_{N-k}} {\left(q^2;q^2\right)_{N-k}}
= q^{2k(b+1)}\cdot \frac{\left(q^{-2b};q^2\right)_{N}}
{\left(q^2;q^2\right)_{N}}\cdot
\frac{\left(q^{-2N};q^2\right)_{k}}
{\left(q^{2(1+b-N)};q^2\right)_{k}}
\end{equation}

\subsection{Elliptic special functions}

The general $\theta$-function with characteristics
is~\cite{Mumford}
$$
\theta_{a,b}(z|\tau) = \sum_{n\in\mathbf{Z}} \mathrm{e}^{\pi i
(n+a)^2\tau}\cdot \mathrm{e}^{2\pi i (n+a)(z+b)}
$$
and we shall use the four standard functions
$$
\theta_{1}(z|\tau) = \theta_{1,1}(z|\tau) = \sum_{n\in\mathbf{Z}}
\mathrm{e}^{\pi i \left(n+\frac{1}{2}\right)^2\tau}\cdot
\mathrm{e}^{2\pi i
\left(n+\frac{1}{2}\right)\left(z+\frac{1}{2}\right)}
$$
$$
\theta_{2}(z|\tau) = \theta_{1,0}(z|\tau) = \sum_{n\in\mathbf{Z}}
\mathrm{e}^{\pi i \left(n+\frac{1}{2}\right)^2\tau}\cdot
\mathrm{e}^{2\pi i \left(n+\frac{1}{2}\right) z }
$$
$$
\theta_{3}(z|\tau) = \theta_{0,0}(z|\tau) = \sum_{n\in\mathbf{Z}}
\mathrm{e}^{\pi i n^2\tau}\cdot \mathrm{e}^{2\pi i n z }
$$
$$
\theta_{4}(z|\tau) = \theta_{0,1}(z|\tau) = \sum_{n\in\mathbf{Z}}
\mathrm{e}^{\pi i n^2 \tau}\cdot \mathrm{e}^{2\pi i
\left(n+\frac{1}{2}\right)\left(z+\frac{1}{2}\right)}
$$
The following identities are used to factorize the Lax matrix and
for the derivation of the defining equations for the operator
$\mathcal{S}_2$
$$
2 \theta_1(x+y)\theta_1(x-y) = \bar\theta_4(x)\bar\theta_3(y)
-\bar\theta_4(y)\bar\theta_3(x)\ ;\ 2 \theta_4(x+y)\theta_4(x-y) =
\bar\theta_4(x)\bar\theta_3(y) +\bar\theta_4(y)\bar\theta_3(x)
$$
$$
2 \theta_2(x+y)\theta_2(x-y) = \bar\theta_3(x)\bar\theta_3(y)
-\bar\theta_4(y)\bar\theta_4(x)\ ;\  2 \theta_3(x+y)\theta_3(x-y)
= \bar\theta_3(x)\bar\theta_3(y) +\bar\theta_4(y)\bar\theta_4(x)
$$
where $\bar\theta_3(z) \equiv \theta_3\left(z |
\frac{\tau}{2}\right)$ and $\bar\theta_4(z) \equiv \theta_4\left(z
| \frac{\tau}{2}\right)$.

The elliptic gamma function~\cite{FV} is defined by the
double product \be \Gamma(z|\tau,\tau^{\prime}) \equiv
\prod_{n,m=0}^{\infty} \frac{1-\mathrm{e}^{2\pi
i\left(\tau(n+1)+\tau^{\prime}(m+1)-z\right)}}{1-\mathrm{e}^{2\pi
i\left(\tau n+\tau^{\prime} m+z\right)}}.\ee
We need the following
properties of this function  \be \Gamma(z+\tau|\tau,\tau^{\prime})
= \mathrm{R}(\tau^{\prime})\cdot \mathrm{e}^{\pi i
z}\theta_1(z|\tau^{\prime})\cdot\Gamma(z|\tau,\tau^{\prime}),\ee
\be \Gamma(z+\tau^{\prime}|\tau,\tau^{\prime})
=\mathrm{R}(\tau)\cdot \mathrm{e}^{\pi i
z}\theta_1(z|\tau)\cdot\Gamma(z|\tau,\tau^{\prime}),\ee where the
constant $\mathrm{R}(\tau)$ does not depend on $z$:
$\mathrm{R}(\tau) = -i\mathrm{e}^{-\frac{\pi i \tau}{4}} \cdot
\left(1;\mathrm{e}^{2\pi i \tau}\right)^{-1}$.

We introduce the function $\gamma(z) \equiv \Gamma(\eta
z|\tau,2\eta)$ which obeys the recurrence relation
\be\label{gamma} \gamma(z+2) = \mathrm{R}(\tau)\cdot
\mathrm{e}^{\pi i \eta z}\ \theta(z)\cdot \gamma(z)\ ;\ \theta(z)
\equiv \theta_1(\eta z|\tau) \ee and can be used to solve the more
general recurrence relation \be\label{solv} \Phi(z+2) =
\frac{\theta(z+a)}{\theta(z+b)} \cdot \Phi(z)\ ; \ \Phi(z) =
\mathrm{e}^{\pi i \eta(b-a)z}\frac{\gamma(z+a)}{\gamma(z+b)},\ee
needed to obtain the operator $\mathcal{S}_2$. There is the useful
formula for the shifted $\gamma$-functions
\begin{equation}\label{transf1}
\frac{\gamma (x-2m)\gamma (y)}{\gamma(x)\gamma(y-2m)} = e^{-\pi i
m\eta(x-y)} \cdot \frac{\left[-\frac{y}{2}+1\right]_{m}}
{\left[-\frac{x}{2}+1\right]_{m}},
\end{equation}
where we use the following notations for the elliptic numbers
\be\label{ellnumbers} [z]=\theta_1(2\eta z)\ ;\ [z]_0=[z]\ ,\
[z]_k =[z]\cdot[z+1]\cdots[z+k-1]\ ,\ k=1,2\cdots \ee To transform
the series with elliptic numbers one uses the transformation
formulae
$$
\frac{\left[A+1\right]_{2k}\left[A+2k\right]_{N-k}}
{\left[A+2k\right]} =
\frac{\left[A\right]_{N}\left[A+N\right]_{k}} {\left[A\right]}\ ;\
\left[B\right]_{2k}\left[B+2k\right]_{N-k} =
\left[B\right]_{N}\left[B+N\right]_{k}\ ;
$$
\begin{equation}\label{transf2}
\frac{\left[-b\right]_{N-k}} {\left[N-k\right]!} =
\frac{\left[1+b-N\right]_{N}\left[-N\right]_{k}}
{\left[-N\right]_{N}\left[1+b-N\right]_k}.
\end{equation}


\begin{thebibliography}{99}

\bibitem{Baxter}
R. Baxter,   
{\it Partition function of the eight-vertex model} \\
 Ann.Phys. NY {\bf 70} (1972),193-228; \\
{\it Eight-vertex model in lattige statistics and
one-dimensional
anisotropic Heisenberg spin chain I. II. III }, 
Ann.Phys , NY {\bf 76} (1973] 1-24, 25-47, 48-71.

\bibitem{FTa}
L.D. Faddeev and L.A. Takhtadzhan  1979 {\it The 
quantum method of the inverse
problem and the Heisenberg XYZ model},  
Russian Math. Surveys 
{\bf 34}, (1979),11-68 (Engl. transl)


\bibitem{KS} P.P. Kulish and E.K.Sklyanin ,
{\it "On the solutions of the Yang-Baxter equation"}
Zap.Nauchn.Sem. LOMI {\bf 95} (1980) 129; \\
{\it "Quantum spectral transform method. Recent developments"},
Lect. Notes in Physics, {\bf v 151}, (1982) , 61


\bibitem{Jimbo}
M.Jimbo ,{\it "Introduction to the Yang-Baxter equation"},
Int.J.Mod.Phys {\bf A 4}, (1983) 3759\\
{\it "Yang-Baxter equation in integrable systems"}, M.Jimbo ed.,
Adv.Ser.Math.Phys., 10 , World Scientific(Singapore) 1990

\bibitem{Faddeev}
L.D. Faddeev,~{\it "How Algebraic Bethe Anstz works
for integrable model"}, Les-Houches lectures 1995, 
hep-th/9605187.

\bibitem{KRS}
P.P. Kulish, N.Yu.Reshetikhin and E.K.Sklyanin, {\it "Yang-Baxter
equation and representation theory"} , Lett.Math.Phys. {\bf 5}
(1981) 393-403

\bibitem{Sklyanin} E.K.Sklyanin,
{\it "On some algebraic structures related to Yang-Baxter
equation"} , Funkz. Analiz i ego Pril. {\bf 16} (1982) pp 27-34; \\
{\it "On some algebraic structures related to Yang-Baxter
equation: representations of the quantum algebra"} ,
Funkz. Analiz i ego Pril. {\bf 17} (1983) pp 34-48\\


\bibitem{BBGK}
  A.~V.~Belitsky, V.~M.~Braun, A.~S.~Gorsky and G.~P.~Korchemsky,
  ``Integrability in QCD and beyond,''
  Int.\ J.\ Mod.\ Phys.\  A {\bf 19} (2004) 4715
  [arXiv:hep-th/0407232].


\bibitem{KZ}
I. Krichever and A. Zabrodin ,~{\it Vacuum curves of elliptic L
operators and Representations of Sklyanin algebra},
arXiv:solv-int/9801022.

\bibitem{Zabrodin}
A.Zabrodin ,~{\it "Commuting difference operators with elliptic
coefficients from Baxter's vacuum vectors"} {\it{J.Phys.A:
Math.Gen.}} {\bf{33}} (2000) 3825

\bibitem{Mumford}
D.Mumford ,~{\it Tata lectures on Theta I}, Prog. in Math.
Birkhauser, 1983

\bibitem{FV}
G.Felder and A.Varchenko, 
  {\it The elliptic gamma function and $SL(3,Z)\times Z^3$},
  arXiv:math.qa/9907061.

\bibitem{FT} I.Frenkel and V.Turaev,
{\it "Elliptic solutions of the Yang-Baxter equation and modular
hypergeometric functions."}, The Arnold-Gelfand Mathematical
Seminars (Cambridge, MA:Birkhauser Boston)(1997) pp 171-204

\bibitem{Rosengren} H. Rosengren,
{\it "Sklyanin invariant integration "}, math.QA/0405072

\bibitem{Gasper}
G.Gasper,{\it "Elementary derivation of summation and
transformation formulas for q-series },
math.CA/9605230 


\bibitem{D}S. Derkachov \textit{Factorization of the
R-matrix. I.}, Zapiski nauchnuch seminarov POMI, v 335 , pp
134-163(2006)[arXiv: math.qa/0503396]\\ \textit{Factorization of
R-matrix and Baxter's Q-operator}, [arXiv:math.qa/0507252].

\bibitem{DM} S. Derkachov, A. Manashov,
\textit{ R-matrix and Baxter Q-operators for the noncompact
SL(N,C) invariant spin chain}, SIGMA 2:084,2006 , nlin.si/0612003


\bibitem{DM1}
  S.~Derkachov and A.~Manashov,
\textit{Baxter operators for the quantum sl(3) invariant spin
chain},
  J.\ Phys.\ A {\bf 39} (2006) 13171 , nlin.si/0604018.

\bibitem{DKK} S. Derkachov , D. Karakhanyan
and R. Kirschner , \textit{Baxter Q-operators of the XXZ chain and
R-matrix factorization}, Nucl.Phys.B738:368-390,2006 ,
hep-th/0511024

\bibitem{KKM}
  D.~Karakhanyan, R.~Kirschner and M.~Mirumyan,
  {\it Universal R operator with deformed conformal symmetry,}
  Nucl.\ Phys.\ B {\bf 636} (2002) 529
  [arXiv:nlin.si/0111032].


\bibitem{DKM}
  S.~E.~Derkachov, G.~P.~Korchemsky and A.~N.~Manashov,
  {\it Noncompact Heisenberg spin magnets from high-energy QCD. I: Baxter
  Q-operator and separation of variables,}
  Nucl.\ Phys.\  B {\bf 617} (2001) 375
  [arXiv:hep-th/0107193].




\end{thebibliography}
\end{document}